\documentclass[%
reprint,
superscriptaddress,
nofootinbib,
amsmath,amssymb,
aps,
prl,
]{revtex4-1}
\usepackage{xr-hyper}
\usepackage{hyperref}
\usepackage{graphicx} 
\usepackage{dcolumn}
\usepackage{bm}
\usepackage{braket}
\usepackage{dsfont}
\usepackage{color,colortbl}
\usepackage{chemmacros}

\usepackage{multirow}
\usepackage{subcaption}
\usepackage{mwe}
\usepackage{booktabs}
\usepackage{caption}
\usepackage{lipsum}
\usepackage{babel,blindtext}
\usepackage{amsmath}
\usepackage[toc,page]{appendix}
\usepackage[symbol*]{footmisc}
\usepackage{float}
\usepackage{natbib}
\usepackage{filecontents}

\makeatletter

\newcommand*{\addFileDependency}[1]{% argument=file name and extension
\typeout{(#1)}% latexmk will find this if $recorder=0
\@addtofilelist{#1}
\IfFileExists{#1}{}{\typeout{No file #1.}}
}\makeatother

\newcommand*{\myexternaldocument}[1]{%
\externaldocument{#1}%
\addFileDependency{#1.tex}%
\addFileDependency{#1.aux}%
}

\myexternaldocument{SI}
\begin{document}
\title{Atomistic evolution of active sites in multi-component heterogeneous catalysts}

\author{Cameron J. Owen$^{\dagger}$}
\affiliation{Department of Chemistry and Chemical Biology, Harvard University, Cambridge, Massachusetts 02138, United States}
\affiliation{John A. Paulson School of Engineering and Applied Sciences, Harvard University, Cambridge, Massachusetts 02138, United States}

\author{Lorenzo Russotto}
\affiliation{Department of Physics, Harvard University, Cambridge, Massachusetts 02138, United States}

\author{Christopher R. O'Connor}
\affiliation{Rowland Institute at Harvard, Harvard University, Cambridge, Massachusetts 02142, United States}

\author{Nicholas Marcella}
\affiliation{Department of Chemistry, University of Illinois, Urbana, Illinois 61801, United States}

\author{Anders Johansson}
\affiliation{John A. Paulson School of Engineering and Applied Sciences, Harvard University, Cambridge, Massachusetts 02138, United States}

\author{Albert Musaelian}
\affiliation{John A. Paulson School of Engineering and Applied Sciences, Harvard University, Cambridge, Massachusetts 02138, United States}

\author{Boris Kozinsky$^{\dagger}$}
\affiliation{John A. Paulson School of Engineering and Applied Sciences, Harvard University, Cambridge, Massachusetts 02138, United States}
\affiliation{Robert Bosch LLC Research and Technology Center, Watertown, Massachusetts 02472, United States}

\def\thefootnote{$\dagger$}\footnotetext{Corresponding authors\\C.J.O., E-mail: \url{cowen@g.harvard.edu}\\B.K., E-mail: \url{bkoz@g.harvard.edu}\\ }\def\thefootnote{\arabic{footnote}}

% custom commands
\newcommand\bvec{\mathbf}
\newcommand{\mathsc}[1]{{\normalfont\textsc{#1}}}

\begin{abstract}
Multi-component metal nanoparticles (NPs) are of paramount importance in the chemical industry, as most processes therein employ heterogeneous catalysts. While these multi-metallic compositions have been shown to result in higher product yields, improved selectivities, and greater stability through catalytic cycling, the structural dynamics of these materials in response to various stimuli (e.g. temperature, adsorbates, etc.) are not understood with atomistic resolution. Here, we present a highly accurate equivariant machine-learned force field (MLFF), constructed from \textit{ab initio} training data collected using Bayesian active learning, that is able to reliably simulate PdAu surfaces and NPs in response to thermal treatment as well as exposure to reactive H$_2$ atmospheres. We thus provide a single model that is able to reliably describe the full space of geometric and chemical complexity for such a heterogeneous catalytic system across single crystals, gas-phase interactions, and NPs reacting with H$_2$, including catalyst degradation and explicit reactivity. Ultimately, we provide direct atomistic evidence that verifies existing experimental hypotheses for bimetallic catalyst deactivation under reaction conditions, namely that Pd preferentially segregates into the Au bulk through aggressive catalytic cycling and that this degradation is site-selective, as well as the reactivity for hydrogen exchange as a function of Pd ensemble size. We demonstrate that understanding of the atomistic evolution of these active sites is of the utmost importance, as it allows for design and control of material structure and corresponding performance, which can be vetted in silico.
\end{abstract}

\maketitle

\subsection*{Introduction}
The lack of atomistic insight into catalyst structure and corresponding performance presents a severely limiting bottleneck in catalyst design. Several examples of these limitations exist, even for monometallic systems, e.g. supported Pt/\ch{CeO2} in the water-gas-shift reaction \cite{Li2021} or Pt under \ch{H2} \cite{Small2014,owen2023unraveling} or CO \cite{Li2021} atmospheres, where active sites have largely been postulated based on chemical intuition, rather than exactly determined with high accuracy and atomic resolution. These challenges are exacerbated when the catalyst geometry or composition increases in complexity, the most trivial extension being an increase in the number of active components, as is the case for multi-component or alloy catalysts. Adding to the increase in configurational complexity in active site idenfitication is the growing evidence that catalyst surfaces become dynamic even under weakly binding adsorbates or mild temperatures, which has been shown for many monometallic (e.g. Pt NPs under \ch{H2} \cite{owen2023unraveling}, Cu surfaces for CO or \ch{NH3} oxidation \cite{Xu_Papanikolaou_Lechner_Je_Somorjai_Salmeron_Mavrikakis_2023}, Au NP surfaces under thermal treatment \cite{Owen2024_au}) and multi-component systems (e.g. \ch{NH3} decomposition on Li-imide surfaces \cite{Yang2023} or PdAg surfaces under thermal treatment \cite{Lim2020a}). Despite added complexity, multi-component heterogeneous catalysts, such as widely studied bimetallic alloy catalysts (e.g. PdAu \cite{Marcella2022,doi:10.1021/acs.jpcc.2c05929}, PdAg \cite{Lim2020EvolutionDynamics}, and PtCu \cite{doi:10.1021/jacs.2c13666}) have garnered increased attention in recent decades due to their demonstrated activity, higher than their bulk counterparts while retaining high selectivity and being less prone to deactivation \cite{doi:10.1021/acscatal.2c00725,doi:10.1021/acs.chemrev.1c00967}. These desirable catalytic properties are typically coupled with low weight loading of the active metal (typically Pt-group or early transition metals) in a noble metal host (e.g. Cu, Ag, or Au). These catalytic performance advantages add to the increased cost-efficiency of the resulting alloy (A$_{x}$B$_{1-x}$, where A is the active metal, B is the host, and x is the atomic weight \%), compared to their single-component alternatives. In this regard, Pd-M, where M is a host metal, and specifically Pd-Au alloy (nanoparticles) NPs have demonstrated significant success in heterogeneous catalysis, particularly in selective oxidation and hydrogenation reactions. However, the atomistic evolution of such catalytic systems have long remained unknown from both experimental and computational perspectives. 

All of these factors are problematic, as the primary focus within the field has increasingly shifted towards rational design of such systems for improved activity and selectivity across a myriad of reactions \cite{doi:10.1021/acscatal.0c03280}, which is implausible without atomistic understanding of how these systems respond to various stimuli encountered during operation. 

Despite careful efforts to understand multi-component single-crystal and NP catalysts across reactive environments via a combination of experimental and computational techniques, atomistic insight into the evolution of the active sites and global catalyst morphologies under realistic conditions remains very limited. This limitation is primarily due to (1) insufficient time- and length-scale experimental resolutions (e.g. slow scanning speeds in scanning tunneling microscopy, lack of atomic resolution in infrared spectroscopy, ensemble averages of X-ray absorption techniques, etc.) for direct observation of dynamic transformations of the system, and (2) the inability of \textit{ab initio} (first principles) methods like density functional theory (DFT) to model appropriate length- and time-scales (\emph{e.g.} number of atoms and simulation time).
Previous studies of bimetallic NP catalyst activity, subject to these limitations, resulted in only tentative hypotheses, rather than definitive conclusions as to what the effects of temperature, `as-synthesized' geometry, or reactive adsorbates are on the evolution of the active site(s) and their subsequent changes in catalytic performance. In the context of these inhibitions, however, attempts have been made to determine the evolution of plausible active sites on PdAu single crystals and NPs as a function of catalyst pretreatment across Pd concentrations. 

For instance, while Pd$_{0.04}$Au$_{0.96}$ NPs embedded in raspberry colloid-templated silica (RCT-SiO$_2$) have shown high activity in the selective hydrogenation of 1-hexyne to 1-hexene \cite{Luneau2020} and remarkable stability, no definitive conclusion was made regarding the cause of such catalytic `success.' Increased activity was attributed to disparate changes in palladium surface content induced by treatment in oxygen, hydrogen, and carbon monoxide at various temperatures, which in turn affected hydrogenation activity, as surface palladium is known to be essential for H$_2$ dissociation and recombination. With the length-scale of Pd active sites as a potentially important descriptor for catalyst performance, several studies have been able to demonstrate the effect of Pd ensemble size for H$_2$ exchange, where the catalyst can effectively operate in dissociative chemisorption- or recombinative desorption-limited regimes \cite{doi:10.1021/acscatal.1c01400}.

Specifically, the work in Ref. \cite{doi:10.1021/acscatal.1c01400} postulated that these regimes are determined by the the size of Pd surface ensembles, e.g. Pd-monomers versus trimers or larger. Hence, small deviations in the concentration of the active metal can lead to drastic changes in reactivity, further complicating the possibility of catalyst design, as such minute changes in the length-scale of the active site have to be understood with atomistic resolution. Outside of the dilute alloy regime, PdAu random alloy NPs at higher Pd concentrations have also shown excellent catalytic activity for heterogeneous molecular formations, outperforming monometallic Pd or Au catalysts \cite{10.1246/cl.200713}. 
Within this higher concentration regime, supported Pd$_{25}$Au$_{75}$ alloy catalysts have been shown to effectively catalyze the reaction of $\alpha,\beta$-unsaturated ketones to silyl enol ethers at room temperature, a feat unobtainable with the pure bulk metal catalysts \cite{doi:10.1021/acscatal.6b02767}. Furthermore, Pd$_{20}$Au$_{80}$ alloy catalysts have been effective in the [2+2+2] cycloaddition of a broad range of alkynes, a method crucial for synthesizing poly-substituted arenes, where the bimetallic catalysts demonstrated efficiency in conditions where monometallic catalysts showed no activity \cite{https://doi.org/10.1002/anie.201800973}. 

Hence, PdAu alloys have demonstrated impressive catalytic performance across compositions for a variety of important chemical transformations, but the active sites responsible for such impressive catalytic activity are not understood in a geometric, and, even more importantly, dynamic sense. 
Intelligent catalyst design and control requires the precise knowledge of the statistical distributions of the active ensemble sizes and geometries, their complementary roles in various reaction steps, and their evolution under reactive and aging conditions. 
In one of the most detailed multi-modal investigations of dilute bimetallic catalysis mechanisms to date, Marcella et al. \cite{Marcella2022} recently employed a combination of X-ray absorption spectroscopy, imaging, and DFT calculations for RCT-SiO$_2$ Pd$_{8}$Au$_{92}$ NPs with a mean size of 4.6 $\pm$ 0.6 nm for active site determination during HD exchange reactions. Following pretreatment with different reactive environments, HD exchange experiments were carried out and coupled with DFT nudged-elastic-band (NEB) calculations for H$_2$ and D$_2$ dissociative chemisorption and recombinative desorption to determine the Sabatier optimum for activity as a function of the size of dilute Pd-ensembles. 
The calculations employed idealized periodic slab models of perfect (111) and stepped (211) $\&$ (331) surfaces to approximate the atomic environments present on the NP surfaces in experiment. 
Pd monomers and dimers were postulated as the active sites controlling the observed catalytic activity after pretreatment of the catalyst in hydrogen, more so than trimers or larger ensembles that form after pretreatment in oxygen. These conclusions were derived from the comparison of the DFT computed barriers at 0 K to experimentally determined apparent activation energies. Like many other preceding investigations, this approach neglected the effects of temperature and dynamics and employed a static assumption of the PdAu surface. The size of Pd ensembles was limited to trimers due to the computational cost needed to accurately study larger ensembles. DFT has also been used to study the surface segregation behavior of these dilute systems under exposure to CO \cite{doi:10.1021/jacs.2c04871}, again limited in the spatial extent of the systems. Hence, despite progress in narrowing down the likely active sites and their contributions to the activity observed in experiments, the critical questions of their precise mechanistic roles and evolution in the course of reaction remain unanswered.

In this work we move past these static assumptions and size limitations of computational models and demonstrate a complete and accurate dynamic study of such important catalyst systems at appropriate length-scales. For this, we develop and apply a machine-learned force field (MLFF), leveraging this method's transformative capabilities and ability to comprehensively describe the complexity associated with multi-component heterogeneous catalysts. In the last few years, machine learning driven molecular dynamics (ML-MD) proved enabling for a variety of challenges within the broad computational materials science field, but its application to heterogeneous catalysts has proven difficult \cite{Bruix2019,Mou2023}. 

\begin{figure*}[!ht]
\includegraphics[width=\textwidth]{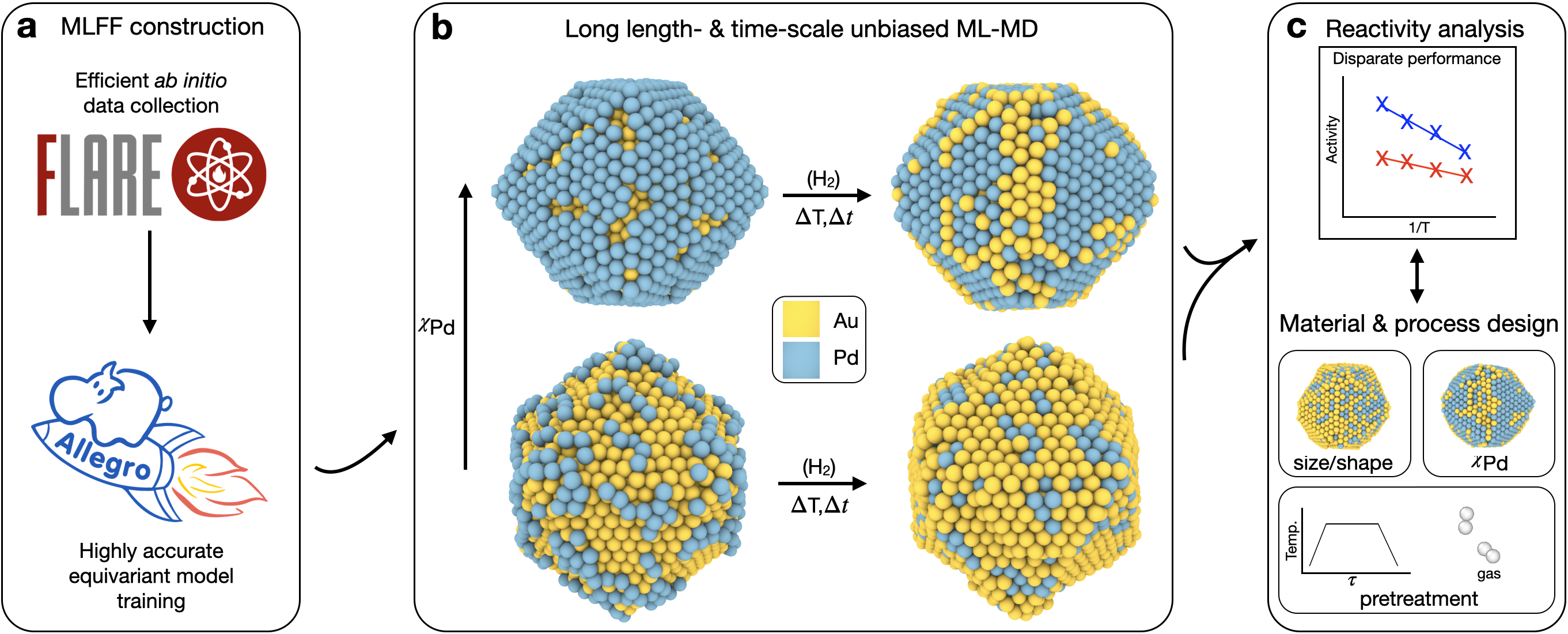}
\caption*{\textbf{Figure 1: Workflow to construct MLFFs for atomistic understanding and control of heterogeneous catalytic systems.} \textbf{a} Combination of FLARE active learning with Allegro MLFF training to yield a flexible model for simulation of bimetallic heterogeneous catalysts. \textbf{b} Once trained, the MLFF enables simulations of a wide selection of nano- and single-crystal catalysts, where both composition and environment can be varied in direct explicit ML-MD simulations. \textbf{c} Catalytic performance is extracted and assessed as a function of catalyst structure and composition, facilitating effective design and control of more active and selective catalysts.}
\label{fig:toc}
\end{figure*}

These difficulties can be assigned to the vast complexity associated with interfacial catalyst systems, where both geometric and chemical space need to be properly sampled in order to generate training data for development of a robust model across such spaces. Critical to the success of these methods is their interpolation across such vast geometric and chemical spaces \cite{Esterhuizen2022}, and also the incorporation of the simulation data to experimental observables \cite{Suvarna2024}, which is the golden standard for the validity of such models. Prior to MLFFs, classical force fields (\emph{e.g.} embedded-atom-method (EAM) \cite{PhysRevB.80.035404} and ReaxFF \cite{Rusalev2022-az}) have been used to describe bulk, alloy, and surface systems of PdAu, but these methods are inherently limited in their ability to describe reactions and large-scale material phenomena like surface reconstructions \cite{Grochola_Russo_Snook_2005} and reactions \cite{Rusalev2022-az} due to their predefined functional forms.

A significant advance in simultaneous accuracy, stability and data efficiency, all critically necessary for describing dynamics of complex reactive systems, were brought forth by equivariant neural network architectures, pioneered by the NequIP MLFF architecture \cite{Batzner2021E3-EquivariantPotentials}, and its descendants Allegro \cite{Musaelian2023}, and MACE \cite{batatia2022designspacee3equivariantatomcentered,NEURIPS2022_4a36c3c5}. Accuracy and stability of models can only be achieved if their training data sets include sufficiently diverse and representative structures. In intricate bimetallic heterogeneous catalysts, the lack of knowledge of mechanisms and which structures to include in training, has been a central obstacle in creating high fidelity ML models. We solve this through a combination of highly efficient active learning in FLARE \cite{Vandermause2022}, which has been demonstrated for H$_{2}$ dissociative chemisorption and recombinative desorption on Pt(111) \cite{Vandermause2022,Johansson2022Micron-scaleLearning}, shape-change of Pt NPs in response to H$_{2}$ exposure \cite{owen2023unraveling}, surface reconstructions of Au \cite{Owen2024_au} and PdAg alloys \cite{Lim2020EvolutionDynamics}, and large-scale dislocations in Cu \cite{owen2024unbiased}.

We combine two of these state-of-the-art approaches, FLARE for the efficient collection of relevant first-principles training data, and Allegro for the development of an equivariant MLFF from these data, ultimately yielding a robust pipeline for the ultimate description of H/PdAu across compositions, chemistries, and configurations. The resulting model is then shown to reliably and accurately predict various properties for these configurationally complex, multi-metallic reactive systems across length-scales, compositions, and reaction conditions. Ultimately, this demonstration serves as a fundamental proof-of-principle for the power of MLFFs to provide atomistic insight into incredibly complex, yet highly influential multi-component catalytic systems across length-scales, compositions, and chemistries with high computational efficiency. A major conclusion from this work is that the dominant active site for heterogeneous reactions can be determined on the fly in our molecular dynamics simulations. Such unbiased, physics-based computational insight opens the door for downstream control and design of catalyst structure and performance, as the effects of chemical synthesis, catalyst pretreatment, and the potential effects of \textit{in operando} degradation, can all be determined \textit{in silico} with quantum mechanical accuracy at experimentally relevant time-/length-scales and conditions.

\subsection*{Results}
\subsection*{Unified MLFF for the full geometric and chemical space for heterogeneous catalysts}
The primary technical challenge underlying this work was to provide a single model that is able to accurately describe the dynamic of reactions and degradation of a multi-component heterogeneous catalyst under reaction conditions. Given such high chemical and structural complexity, we employed active learning in the FLARE framework to collect ab initio data in an efficient manner, the frames of which were then used to train an Allegro MLFF, as shown in Fig. 1a. These techniques are described in more detail in the Methods section. Ultimately, we provide a framework by which robust MLFFs can be constructed for multi-component catalysts using the combination of FLARE \cite{Vandermause2022} and Allegro \cite{Musaelian2023}. 

\begin{figure*}
\centering
\includegraphics[width=\textwidth]{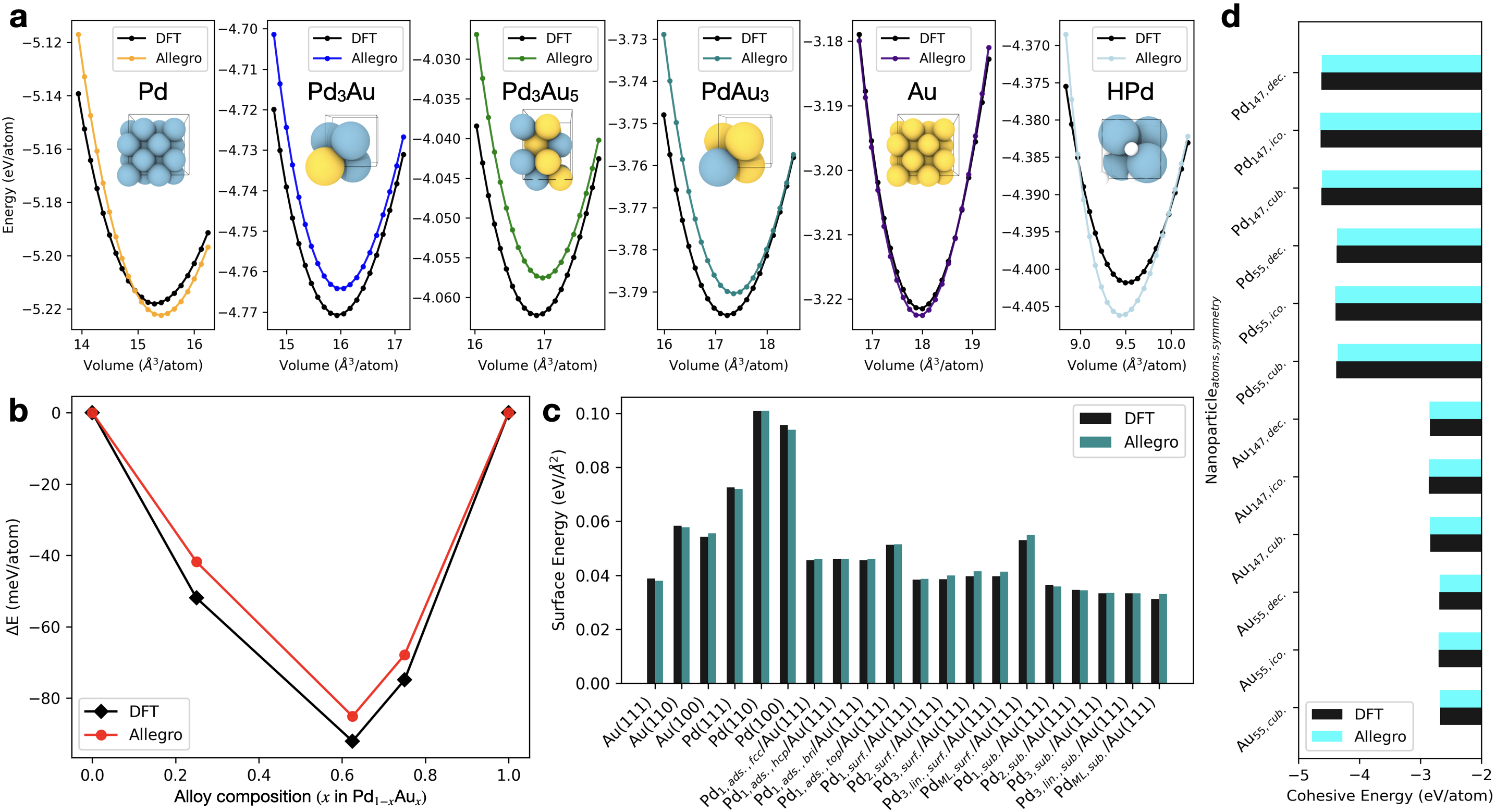}
\caption*{\textbf{Figure 2: Validation of the Allegro MLFF for bulk, surface and nanoparticle alloys.} \textbf{a} Energy (eV per atom) versus volume (\AA{}$^3$ per atom) curves for bulk crystalline Pd, Pd$_3$Au, Pd$_3$Au$_5$, PdAu$_3$, Au, and HPd, respectively, as predicted by DFT (black) and Allegro (colored). \textbf{b} Convex hull for the Pd$_{1-x}$Au$_x$ composition alloys comparing the $\Delta E$ (meV per atom) predicted by DFT (black) and Allegro (red). \textbf{c} Surface energy (eV \AA{}$^{-2}$) for a variety of Pd/Au monometallic and Pd-Au alloy surfaces as predicted by DFT (black) and Allegro (teal). \textbf{d} NP cohesive energies predicted by DFT (black) and Allegro (light blue) for a range of NP sizes and shapes.}
\label{fig:bulk_surf_val}
\end{figure*}

\subsection*{Allegro matches DFT across a broad range of compositions, configurations, and chemistries}
\subsection*{Au-Pd bulk-alloys, hydrides, and surfaces}
We then trained an equivariant Allegro MLFF on the complete set of first-principles data collected using FLARE active learning as described in the previous section. During this process, we explicitly tested the effect of varying model architectures on a wide-ranging set of validation targets. Ultimately, a model with an angular resolution of $\ell_{max}=2$ and symmetric pairwise cutoff matrix with values 6.0, 5.0, and 5.0 \AA{} was employed for the Au-Au/Pd-Pd, H-Au/H-Pd, and H-H interactions, respectively. These parameters appeared to be most critical for this system, and all others are discussed in detail in the Methods section. The choice of a larger cutoff for the H-H interaction in the larger cutoff matrix was found to be necessary for a more accurate description of bulk Pd-hydride, since smaller cutoffs resulted in a drastic shift of the predicted minimum in the bulk energy as a function of volume to lower, nonphysical values. 

Following training, the Allegro MLFF was then independently tested against DFT for prediction of properties across the entire composition range for bulk, surfaces, and NPs containing various mixtures of H, Pd, and Au. These results are provided in part in Fig. 2. In panel \textbf{a}, bulk energy-volume curves are shown across the entire set of stable configurations in the binary alloy composition space, as well as the lowest energy phase for Pd hydride. By then taking the minimum of each of these curves predicted by both DFT or Allegro, we computed the energy of formation for each alloy composition, constructing the convex hulls, shown in panel \textbf{b}. Excellent agreement is observed between DFT and Allegro, for all targets, with errors in the energies of formation below 10 meV/atom. 

\begin{figure*}
\centering
\includegraphics[width=\textwidth]{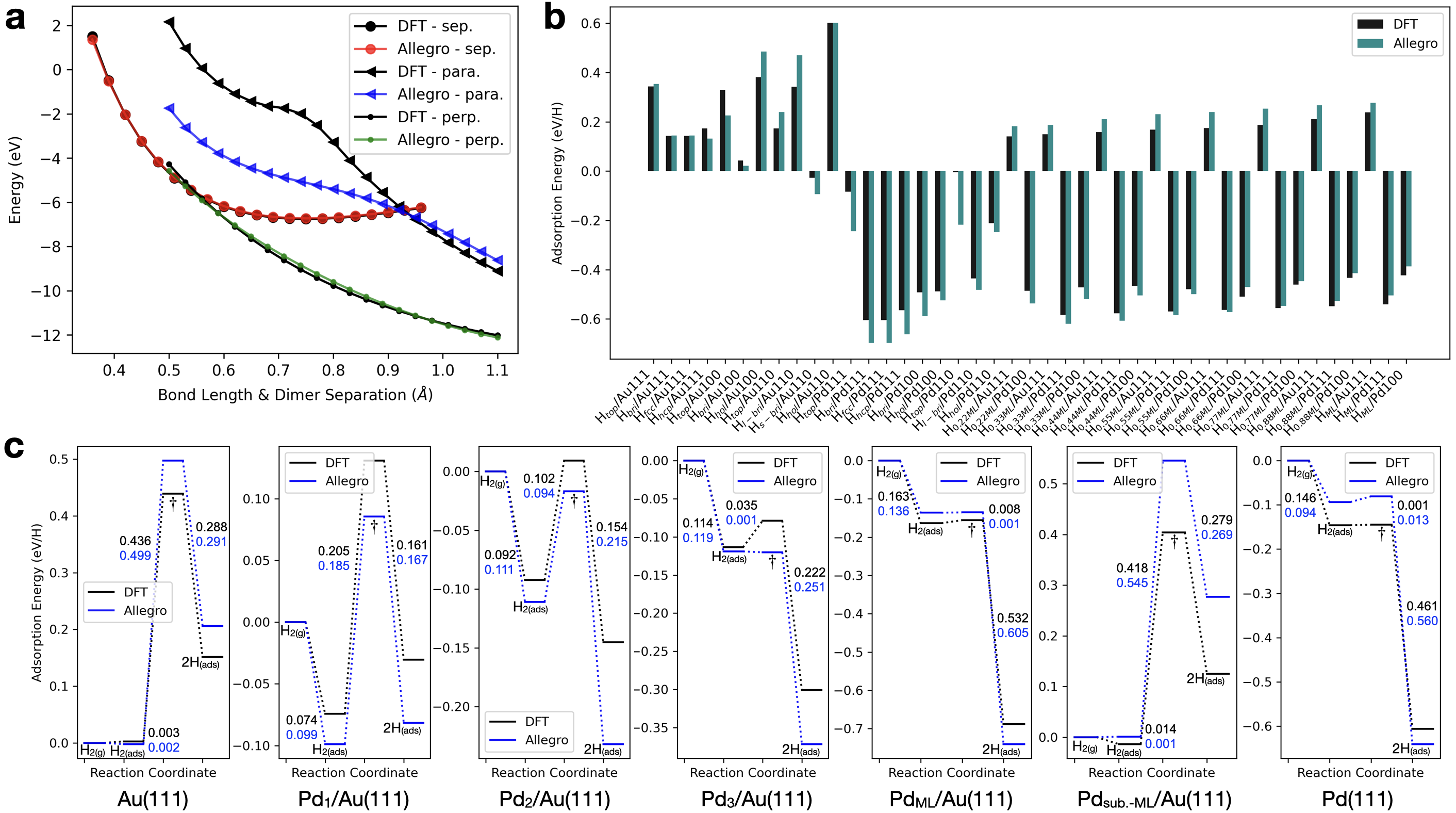}
\caption*{\textbf{Figure 3: Adsorbate and adsorption validation of the MLFF against DFT.} \textbf{a} H$_2$ separation and dimer interaction energies (eV) predicted by DFT (black) and Allegro (red, blue, and green, respectively).  \textbf{b} H adsorption energies (eV/H) on various Au and Pd surfaces and across H coverage predicted by DFT (black) and Allegro (teal). \textbf{c} H$_2$ physisorption $\to$ 2H chemisorption reaction barriers for pure Au(111), dilute Pd ensembles in Au(111), and Pd(111) predicted by DFT (black) and Allegro (blue). Transition states are identified by a cross.}
\label{fig:ads}
\end{figure*}

In addition to the bulk alloy structures, we also considered the prediction of surface energies and nanoparticle cohesive energies across a wide range of configurations and compositions, the alloys being constrained to the (111) facet of Au(111), while the NPs were evaluated across a variety of particle sizes and shapes. These results are provided in panel \textbf{c} and \textbf{d}, respectively, where excellent agreement is again observed between Allegro and DFT across all surface facets and NP sizes and shapes, including variations in composition for the surface facets. Importantly, the MLFF surface energy predictions show reliable transferability across various dilute alloy structures. Consequently, the segregation energy of Pd, defined as
\begin{equation}
    E_{seg.,Pd_{1}} = E_{Pd_{1},surf.} - E_{Pd_{1},sub-surf.},
\end{equation}
, has an error of only 2.56 meV per atom and thus correctly captures the energetic preference for isolated Pd to surround itself with Au by segregating into the subsurface layer. As for the NPs in panel \textbf{d}, high accuracy is again obtained for the prediction of cohesive energies across sizes and shapes. Notably, while the Allegro training set only included particles with icosahedral and cuboctahedral symmetries, the MLFF model accurately describes the ino-truncated decahedral symmetry, with lower than 1 meV/atom energy errors. Hence, these results provide confidence in the combined FLARE and Allegro MLFF training workflow for the prediction of bulk, surface, and NP properties across the complete range of alloy compositions and to relevant structures like Pd-hydrides which form under long time-scale exposure of these systems to reactive H$_2$ atmospheres.

\subsection*{MLFF reliably predicts adsorption energies across catalyst concentrations}
Extending from the bulk, surface, and NP evaluations provided in the previous section, we then evaluated the Allegro MLFF on descriptions of gaseous interactions of H, H$_2$, and both of these species adsorbed on a variety of monometallic and alloyed PdAu surfaces. Moreover, we recomputed the various pathways provided in the supplemental information of Ref. \cite{Marcella2022} at the DFT level of theory employed for the MLFF trained here for H$_2$ dissociative chemisorption and recombinative desorption on the monometallic and dilute alloy structures from pure Au(111) to Pd(111), while the Au(211), and Au(331) to Pd(211), and Pd(331) surfaces are provided in Suppl. Fig. 1. These reaction barrier data follow the given profile,
\begin{equation}
    \ch{H2 <>[ $k_{\mathrm{ads.}}$ ][ $k_{\mathrm{des.}}$ ] H2$^*$ <>[ $k_{\mathrm{diss.}}$ ][ $k_{\mathrm{reco.}}$ ] }2\textrm{H}^*
\end{equation}
where \ch{H2} is the gas-phase reactant, $k_{\mathrm{ads.}}$ is the rate for \ch{H2} physisorption (denoted as \ch{H2$^*$}), $k_{\mathrm{des.}}$ is the rate for \ch{H2} desorption of the physisorbed state, $k_{\mathrm{diss.}}$ is the rate for dissociative chemisorption to enter the 2H$^*$ state, and $k_{\mathrm{reco.}}$ is the rate for recominative desorption to form \ch{H2$^*$} and enter back into the gas-phase as \ch{H2}. Moreover, the chemisorbed 2H$^*$ atoms are able to diffuse over the surface via spillover mechanisms from Pd ensembles to Au and back to Pd for recombination. The complete set of results is provided in Fig. 3. In panel \textbf{a}, the interaction energies between isolated gas-phase H and itself (H$_2$), as well as between H$_2$ molecules are both accurately described by the Allegro model. The interaction of H$_2$ dimers in a parallel orientation (blue markers) appears to agree less favorably with the DFT computed labels compared to the other orientations, but we note that the qualitative agreement remains. 

\begin{figure*}
\includegraphics[width=1.0\textwidth]{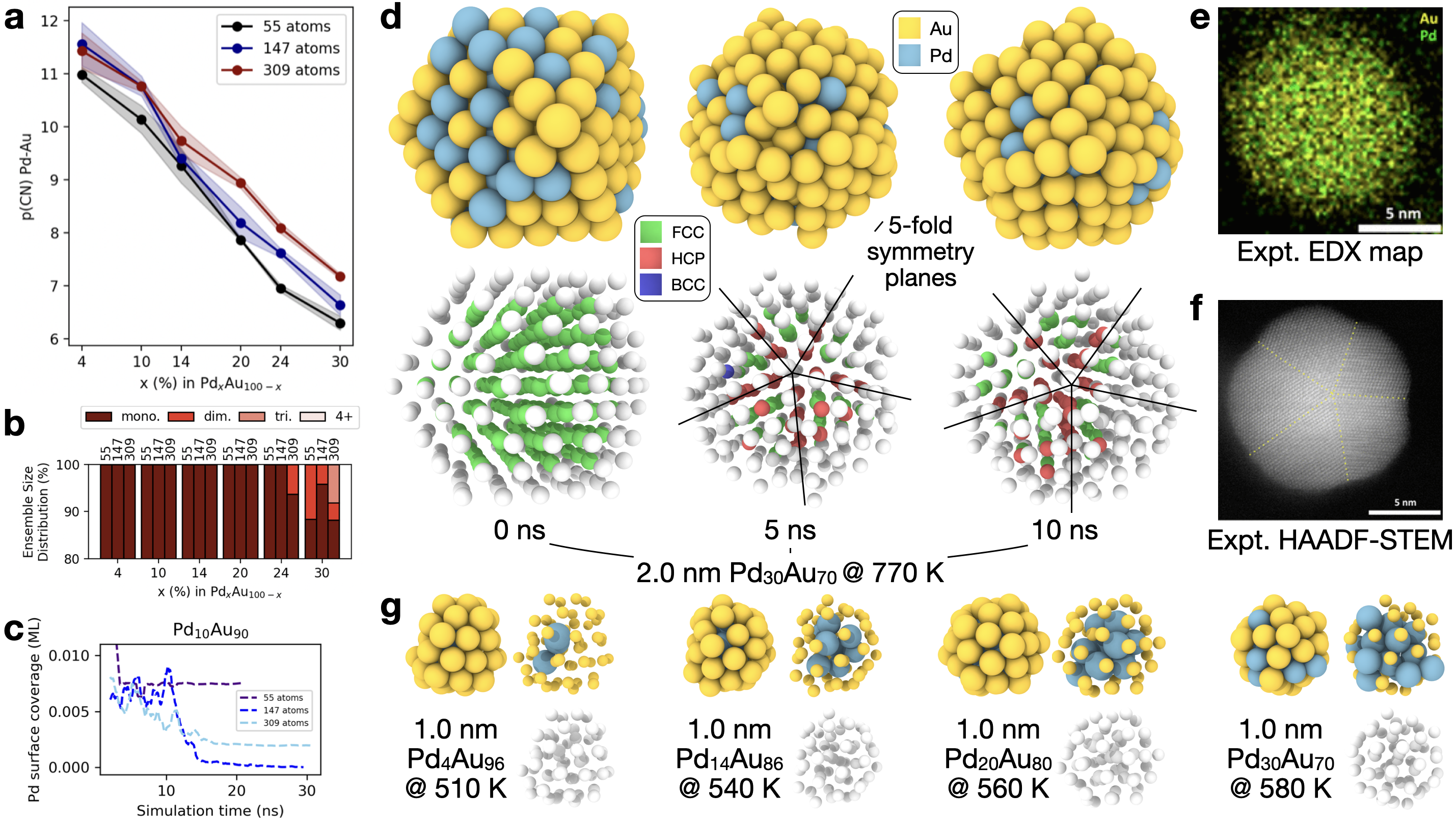}
\caption*{\textbf{Figure 4: Allegro MLFF for NP evolution predicts experimental symmetries and active site distributions.} \textbf{a} Pd-Au partial coordination number computed from ML-MD simulations for particles of varying alloy composition and size. Shaded regions represent the standard deviation amongst three simulations for each system with random alloying seeds. \textbf{b} Surface Pd-ensemble size distribution, as a function of NP size and grouped by alloy composition. Columns represent the size distribution during the final quenching phase, showing that monomers dominate the distribution. \textbf{c} Coverage evolution as a function of particle size and simulation time for a composition of Pd$_{10}$Au$_{90}$. The graph provides the amount of Pd in the surface of the NP, expressed as coverage in monolayers (ML), as a function of simulation time (in ns). \textbf{d} Evolution of a 309 atom particle ($\approx$ 2 nm diameter) with 30\% Pd randomly substituted into the Au lattice at t = 0 ns. Simulation at 770 K for 10 ns shows appearance of pentatwinned symmetry and only small length-scale Pd ensembles at the surface (e.g. monomers and dimers). This global symmetry was not captured during MLFF training but the resulting model is able to predict its emergence, in accordance with experimental observations, which is important for catalysis as the pentatwin boundaries influence the surface termination of atoms, thus influencing catalytic activity \cite{song2024uneven}. The symmetry planes are denoted by black line segments, and are identified using polyhedral template matching in Ovito \cite{ovito}. \textbf{e} Experimental energy dispersive X-ray map, adapted from Ref. \cite{vanderHoeven2021EntropicCatalysts} to demonstrate the presence of rather uniform alloying of Pd into gold after annealing. \textbf{f} Experimental high-angle annular dark-field imaging scanning transmission electron microscopy image of a particle like the one in \textbf{e} to demonstrate the presence of pentatwinned symmetry. \textbf{g} Similar particle evolution to that shown in \textbf{c} but with a smaller 55 atom NP (1.0 nm in diameter) with varying amounts of Pd content to establish the effect of size on alloy structures observed. Snapshots for the full NP, NP with reduced radius for Au atoms, and PTM analysis of each particle are provided, along with the size, compositions, and annealing temperatures.}
\label{fig:mix}
\end{figure*}

\begin{figure*}
\centering
\includegraphics[width=\textwidth]{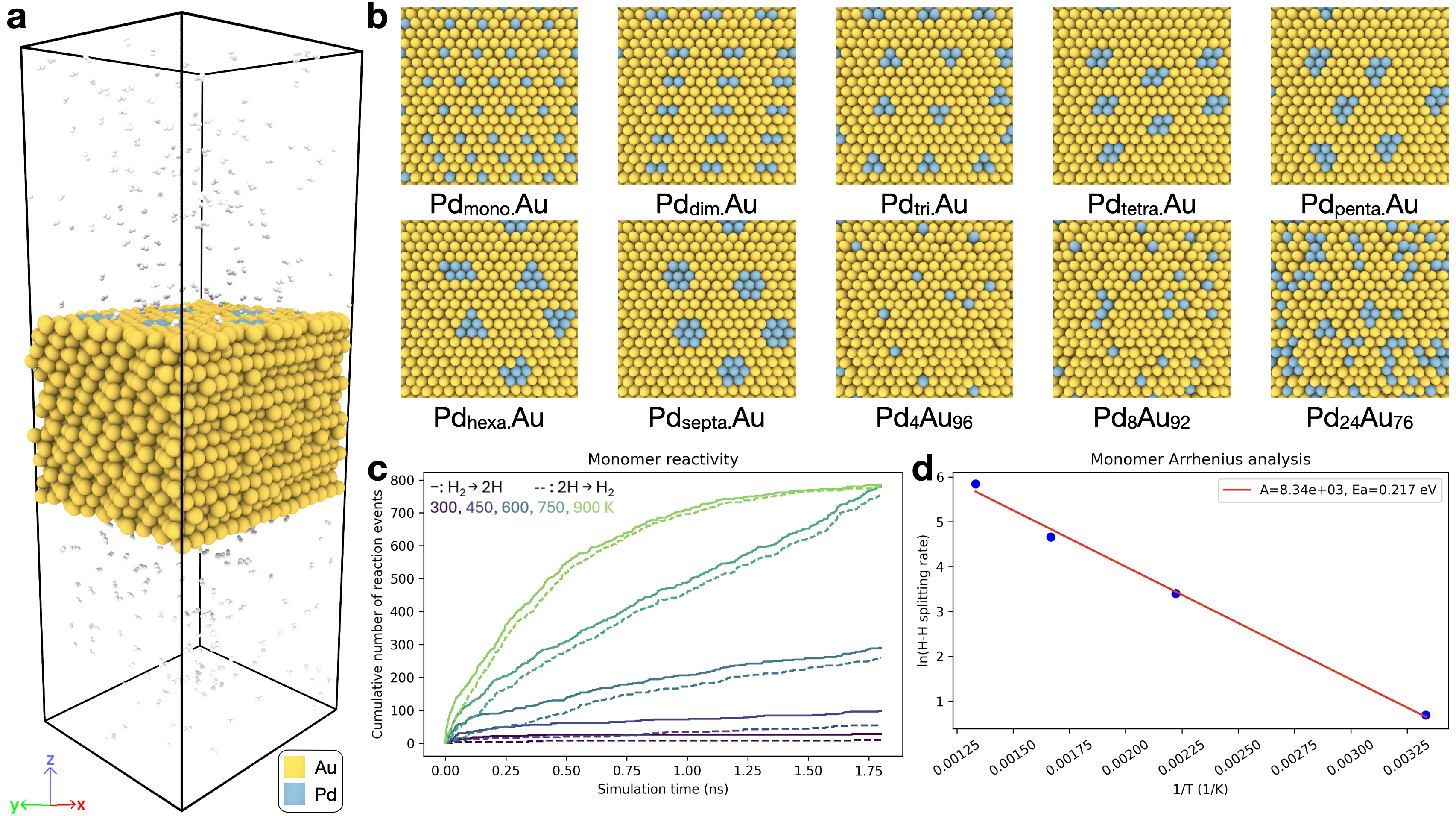}
\caption*{\textbf{Figure 5: Atomistic reactivity and active site determination made possible.} \textbf{a} Catalytically active nano-cell (4526 atom) employed for all reactivity analyses. Au atoms are gold, Pd are green, and H are white. \textbf{b} Snapshots of the various surface compositions studied, from explicitly constructed Pd-monomers to septamers, as well as randomly alloyed 4 \%, 8 \%, and 24 \% Pd systems. \textbf{c} An example of the reactivity assessment performed for each system, here performed for the explicitly constructed Pd-monomer system, where the cumulative number of dissociative chemisorption and recombinative desorption events are tracked as a function of simulation time and temperature. \textbf{d} Example Arrhenius analysis, performed using the last 1 ns of data from panel \textbf{c}, where the natural log of the H-H exchange rate is plotted against inverse temperature. These data are then fit using linear regression to determine the Arrhenius pre-exponential factor and apparent activation energy for the reaction over Pd-monomers.}
\label{fig:react}
\end{figure*}
To ensure the stability of molecular dynamics, it is sufficient that molecules strongly repel each other at close distances, while high accuracy is not necessary for the high energies of those configurations, which are very unlikely to be encountered in the production ML-MD runs. Panel \textbf{b} displays adsorption energy predictions for chemisorbed H atoms from dilute to full monolayer coverages as predicted by both DFT and the Allegro model on pure Au and Pd surfaces. Despite small shifts in the absolute values of the energy predictions, the relative energies between adsorption sites, especially with respect to the surface facet, as well as increasing coverage remain in good agreement with the DFT computed labels, with total energy errors never exceeding 2 meV/atom across the entire set.

We computed reaction barriers for H$_2$ dissociative chemisorption and recombinative desorption on (111), (211), and (331) facets of Au, Pd, and intermediate alloy structures in panels \textbf{c}, and Suppl. Fig. 1, respectively, ranging from pure Au to pure Pd. These explicit tests are directly relevant to the catalytic pathways on dilute alloys of PdAu, where small length-scale Pd ensembles are posited to dominate the overall activity. Similarly to panel \textbf{b}, the relative energies of all initial physisorption, transition, and final chemisorption states are accurately predicted by the Allegro model, with total energy errors below 2 meV/atom. Importantly, all of the reaction energy profiles are consistent between Allegro and DFT. Hence, the MLFF is exhaustively shown to provide excellent predictions for the reactions and relative energies of bulk, surfaces, NPs, and alloy compositions, providing confidence in the robustness of the model used in ML-MD simulations in the following sections.

As an important first demonstration of the MLFF in long length- and time-scale ML-MD, we simulated the effect of annealing for small NPs (with diameters 1-2 nm) and studied the appearance of alloy structures as a function of size and composition, as the resulting Pd distribution and effects of local Pd concentration hold enormous importance in heterogeneous catalytic processes. Hence, a flexible protocol for studying the appearance of catalytically relevant active site structures as according to the underlying alloy behavior of such complicated systems \textit{in silico} is provided.

\subsection*{Annealing allows for Pd redistribution in NPs}
Here we focus on NP annealing, which is a nearly universal step in the catalyst pretreatment process in order to remove contaminants remaining from catalyst synthesis using either wet or dry methods. Importantly, our simulations directly predict the amount and distribution of Pd on the surface, which is paramount for understanding subsequent activity and selectivity of the system given the amount and geometry of the active site ensembles. We demonstrate in Fig. 4 the structural evolution of 1-2 nm NPs across the experimentally relevant composition range from 4 to 30 \% Pd content. Panel \textbf{a} shows an analysis of the partial coordination number (number of Au atoms in the first coordination shell of Pd) for 3 different nanoparticle sizes across the range of compositions that are thermally equilibrated. All particle sizes behave similarly, exhibiting a monotonic decreasing trend in the Pd partial coordination number as particle size increases. This observation is directly in line with what would be observed from ensemble averages collected using X-ray absorption spectroscopy, where effectively no differences are observed across these samples. We then asked the question, where is the Pd going over the course of each annealing simulation, the details of which are provided in the Methods section, and what sort of surface ensembles are presenting themselves throughout annealing of the NP. Briefly, the particles begin in a melted configuration of only Au atoms, Pd is randomly substituted up to the listed concentration, then the particles are heated slowly (20 K per ns) to their respective melting temperatures, and quenched back down to 300 K using the same heating rate to observe structural preferences as a function of concentration and size. 

In panel \textbf{b}, we show the Pd surface ensemble size distribution at the end of the quenching phase to gain insight into the final distribution of active sites ensemble size. In panel \textbf{c}, we fix the composition at Pd$_{10}$Au$_{90}$, but vary the NP size and track how much Pd remains in the surface layer as a function of simulation time. We find that Pd surface coverage (in units of monolayers), decreases from the initially mixed state and approaches very small amounts as the simulations progress. Some interesting observations can be made here: (1) there is a non-monotonic trend in the amount of surface Pd relative to the NP size and (2) upon quenching, the exposed Pd on the surface is found predominantly in the form of isolated monomers. More generally, monomers make up 100\% of the surface ensembles for all NP sizes for the lower Pd concentration alloys. 

An important aspect of alloying is that the global symmetry of the NP can change. Hence, we demonstrated that the MLFF can also predict experimentally relevant particle morphologies, as shown in Fig. 4\textbf{d}-\textbf{f}, where we focus on a 309 atom Pd$_{30}$Au$_{70}$ particle as evolved during steady state conditions in the same simulations that were analyzed in Fig. 4\textbf{a}-\textbf{c}. We can observe the reduction in Pd content in the surface of the particle, as well as the decrease in length-scale of the ensembles, but more importantly, we see the appearance of a global symmetry change in the particle. Ultimately, we observe pentatwinned symmetry, the planes being denoted by the black lines in panel \textbf{d}, assigned to the HCP planes provided by the polyhedral template analysis performed in Ovito \cite{ovito}, the structure of which has been observed experimentally using HAADF-STEM as in panel \textbf{f} for a larger particle ($\approx$ 10 nm in diameter) with lower Pd concentration ($\approx$ 4 \%) from Ref. \cite{vanderHoeven2021EntropicCatalysts}. We can also qualitatively confirm the amount of mixing by comparing our partial coordination numbers and visual analyses to the experimental EDX map in panel \textbf{e}, as there is no agglomeration of Pd evidenced by the uniformity of the 2D projection of the atomic positions in green. These are very important findings in this work, as both mixing and morphology can be accurately captured using our method, leading to robust determination of particle faceting as a consequence of the global symmetry, as well as the size and amount of active sites on such facets.

Lastly, we also provide a series of snapshots for the smallest particles, where despite the small particle diameter, the model still predicts preferential segregation of Pd into the core, which results in little to no accessible Pd for catalysis unless the Pd concentration is increased. This has been demonstrated by Ricciardulli et al. \cite{doi:10.1021/jacs.1c00539}, where segregation of Pd influenced the reaction barrier for hydrogen production, as smaller Pd ensembles resulted in a destabilizing effect of the transition states for O-O cleavage, meaning that the catalytic reduction of \ch{O2} to \ch{H2O2} became more selective towards hydrogen peroxide formation due to the change in length-scale of the active Pd ensembles, but with reduced activity due to the reduction of total Pd content in the surface of the 10 nm particles. These results mean that our MLFF is able to provide direct atomistic insight into the types of ensembles that are present on NP alloy catalysts, leveraging its ability to describe both the reaction potential energy surface and alloy behavior for these multi-component NPs under relevant conditions.

\subsection*{MLFF reliably predicts reaction complexity across alloy space}
Given the alloying results provided in the previous section, where small length-scale Pd ensembles are shown to be the prevalent species following annealing of dilute PdAu alloy nanoparticles, we then tasked our MLFF with predicting their reactivity for hydrogen exchange as a function of their size.
Hence, we tasked our MLFF to describe a variety of these alloy systems under reactive hydrogen atmospheres to directly observe the active site selectivity using reactive ML-MD simulations. In order to compare to experimental results of larger NPs ($>$ 5 nm) and single crystals, we employed a catalytically active nano-cell (4,526 atoms) comprised of an alloy slab with varying composition, from 4-24 \% Pd in the near surface region (top two atomic layers). This composition range exactly corresponds to that for which experimental measurements for HD exchange are available. The specific details pertaining to these simulations are provided in the Methods section.

The initial set of results for reactivity are provided in Figure 5, where the effects of composition, temperature, and hydrogen pressure are all varied to study catalyst activity and stability. The nano-cell employed for all reactivity and mechanistic insights pertaining to active site ensembles is shown in panel \textbf{a}, where Au (gold), Pd (blue-gray), and H (white) atoms can be observed. This same nano-cell was employed for all reactivity measurements across ensemble sizes of Pd, where the Pd was introduced either explicitly or randomly into the top two surface layers of both sides of the slab when initializing the ML-MD simulation. Substitution of the Pd composition into the full Au host, rather than just the surface layers of the system, was not done in order to prohibit large deviations in the lattice spacing (as the lattice parameter of Pd is shorter than Au), which would permit residual effects of unintentional mechanical strain.

\begin{figure}
\centering
\includegraphics[width=\columnwidth]{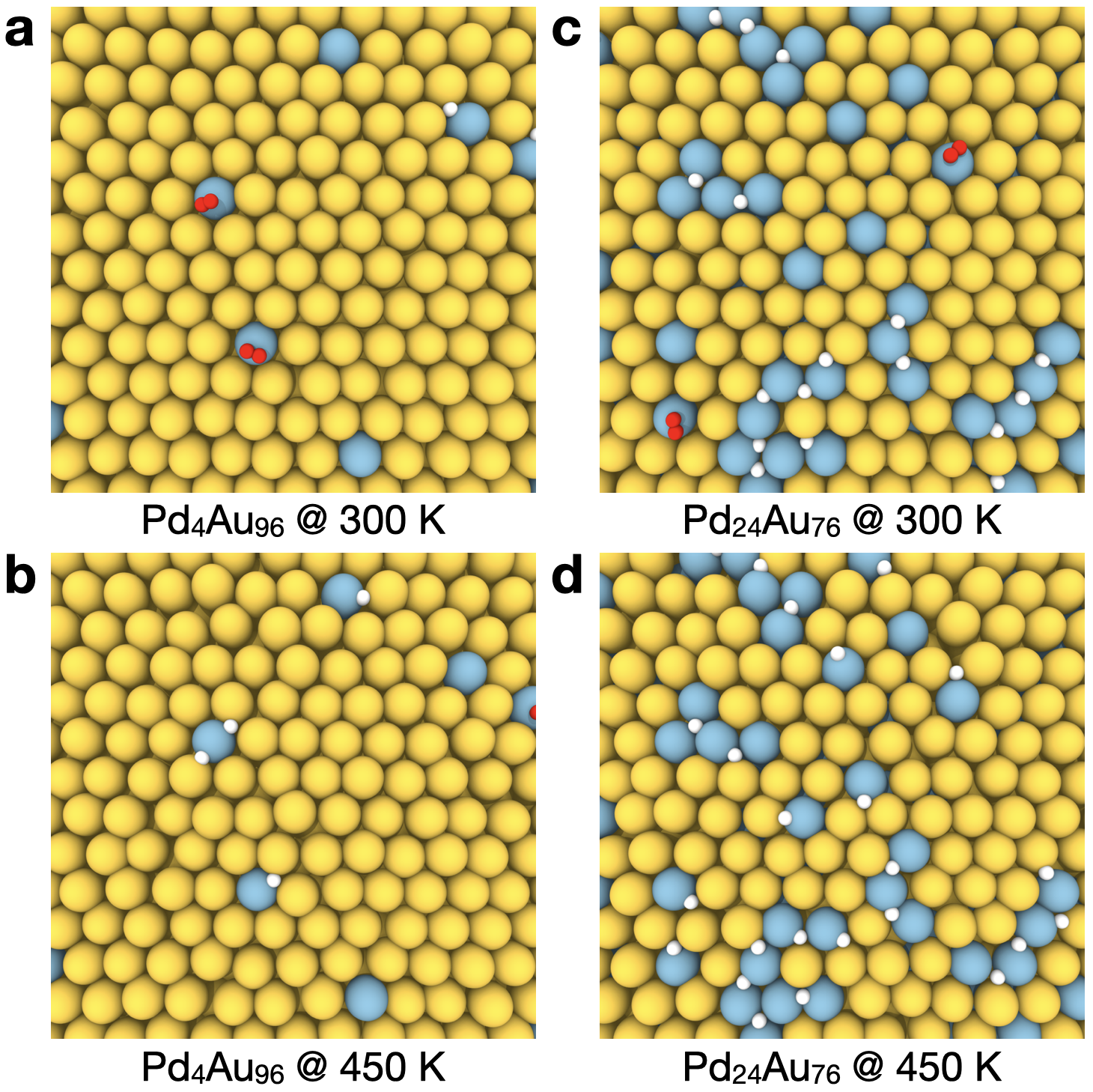}
\caption*{\textbf{Figure 6: Simulations capture temperature dependence of non-reactive physisorption and dissociative chemisorption.} \textbf{a} Snapshot of the Pd$_{4}$Au$_{96}$ catalyst at 1 ns under a thermostat of 300 K. Physisorbed hyrogen dimers are colored in red, and chemisorbed hydrogen atoms are colored white. \textbf{b} The same catalyst as in \textbf{a} but at 450 K. The increase in temperature leads to an increase in the reactivity of Pd-monomers, as fewer physisorbed molecules are observed. \textbf{c} Snapshot of the Pd$_{24}$Au$_{76}$ catalyst at 1 ns under a thermostat of 300 K. Here, large Pd-ensembles are fully decorated by chemisorbed H, whereas only physisorption of H$_2$ is observed on the monomers. \textbf{d} The same catalyst as in \textbf{c}, but at 450 K. Again, the monomers become active, and no physisorption is observed.}
\label{fig:phys}
\end{figure}

\begin{table}[!htbp]
\centering
\resizebox{\columnwidth}{!}{\begin{tabular}{|c|c|c|c|c|}
\hline
\bf{} & \multicolumn{2}{|c|}{\bf{Dissociative chemisorption}} & \multicolumn{2}{|c|}{\bf{Recombinative desorption}} \\
\hline
\bf{System}  & \bf{E$_{\text{act}}$ (eV)} & \bf{Pre-exp. factor (ns$^{-1}$)} & \bf{E$_{\text{act}}$ (eV)} & \bf{Pre-exp. factor (ns$^{-1}$)} \\
\hline
Pd$^{(1)}$Au & 0.233 & 1.13E13 & 0.240 & 1.29E13 \\
Pd$^{(2)}$Au & 0.339 & 7.42E13 & 0.297 & 3.70E13 \\ 
Pd$^{(3)}$Au & 0.332 & 6.30E13 & 0.333 & 6.60E13 \\ 
Pd$^{(4)}$Au & 0.333 & 4.83E13 & 0.429 & 2.51E14 \\ 
Pd$^{(5)}$Au & 0.388 & 1.52E14 & 0.363$^*$ & 9.93E13$^*$ \\ 
Pd$^{(6)}$Au & 0.376 & 9.53E13 & 0.418 & 1.96E14 \\ 
Pd$^{(7)}$Au & 0.373 & 8.64E13 & 0.413 & 1.77E14 \\ 
\hline
Pd$_{4}$Au$_{96}$ & 0.162 & 6.35E11 & 0.196 & 1.19E12 \\
Pd$_{8}$Au$_{92}$ & 0.203 & 2.30E12 & 0.218 & 2.86E12 \\
Pd$_{24}$Au$_{76}$ & 0.285 & 3.86E13 & 0.278 & 3.43E13 \\
High P. - Pd$_{24}$Au$_{76}$ & 0.254 & 2.77E13 & 0.275 & 3.86E13 \\
Pd$_{100}$Au$_{0}$ & 0.370 & 1.72E14 & 0.524 & 1.52E15 \\
\hline
\hline
\bf{Expt. System} & \multicolumn{2}{|c|}{\bf{E$_{\text{act}}$ (eV)}} & \multicolumn{2}{|c|}{\bf{Pre-exp. factor ($\mu$mol$_{\textrm{HD}}$s$^{-1}$g$^{-1}$)}} \\
\hline
Pd$_{4}$Au$_{96}$ \cite{vanderHoeven2021EntropicCatalysts} & \multicolumn{2}{|c|}{0.33$\pm$0.06} &  \multicolumn{2}{|c|}{2.1$\times$10$^8$}  \\
Pd$_{8}$Au$_{92}$ \cite{vanderHoeven2021EntropicCatalysts} & \multicolumn{2}{|c|}{0.59$\pm$0.07} &  \multicolumn{2}{|c|}{3.1$\times$10$^{12}$}  \\
Pd$_{100}$Au$_{0}$ \cite{vanderHoeven2021EntropicCatalysts} & \multicolumn{2}{|c|}{0.46$\pm$0.01} &  \multicolumn{2}{|c|}{1.8$\times$10$^{12}$}  \\
\hline
\end{tabular}}
\caption*{\textbf{Table 2: Activation energies and pre-exponential factors determined from simulation.} Activation energies for both dissociative chemisorption and recombinative desorption, as well as their pre-exponential factors, as determined from Arrhenius analysis from the ML-MD simulations using data collected across 450, 600, and 750 K. Data from 300 and 900 K were excluded given the former being unreactive, and the latter deactivating. $^{*}$ Arrhenius fits only possible using 600 and 750 K data, as no reactions occurred in the 450 K data over the last 1 ns of simulation. Experimental data from Ref. \cite{vanderHoeven2021EntropicCatalysts} collected at a pressure of 1 atmosphere. \label{tab:act}}
\end{table}

\subsection*{Hydrogen exchange is dependent on Pd ensemble size}
In panel \textbf{b} of Fig. 5, we provide snapshots of all of the surface compositions we explored under the effects of hydrogen and annealing. For all explicitly constructed Pd ensembles of different length scales, we constrained the total number of Pd across both surfaces to be $\approx$ 60 atoms. The last three systems, specifically Pd$_{4}$Au$_{96}$, Pd$_{8}$Au$_{92}$, and Pd$_{24}$Au$_{76}$ were constructed via random substitutions of Au with Pd in the top two surface layers of the catalyst. These systems were then evolved with and without the presence of a hydrogen atmosphere, while performing surface reaction counting (e.g. number of \ch{H2} dissociation and recombination events) and analyzing the Pd ensemble geometries. An example of the reaction counting is provided in panel \textbf{c} of Fig. 5, where cumulative dissociative chemisorption and recombinative desorption events were logged as a function of simulation time and temperature. These data were obtained by tracking changes in the coordination number of H atoms with other H atoms, where a change from 0 to 1 denoted a recombination event, and 1 to 0 denoted a splitting event. These data are then used for Arrhenius analyses using linear regression across the final 1 ns of data for each temperature and each catalyst composition. As can be seen in the plot for monomers, the reaction rate at 900 K is not constant, but is decreasing with time, which denotes catalyst deactivation. 

\begin{figure*}
\centering
\includegraphics[width=\textwidth]{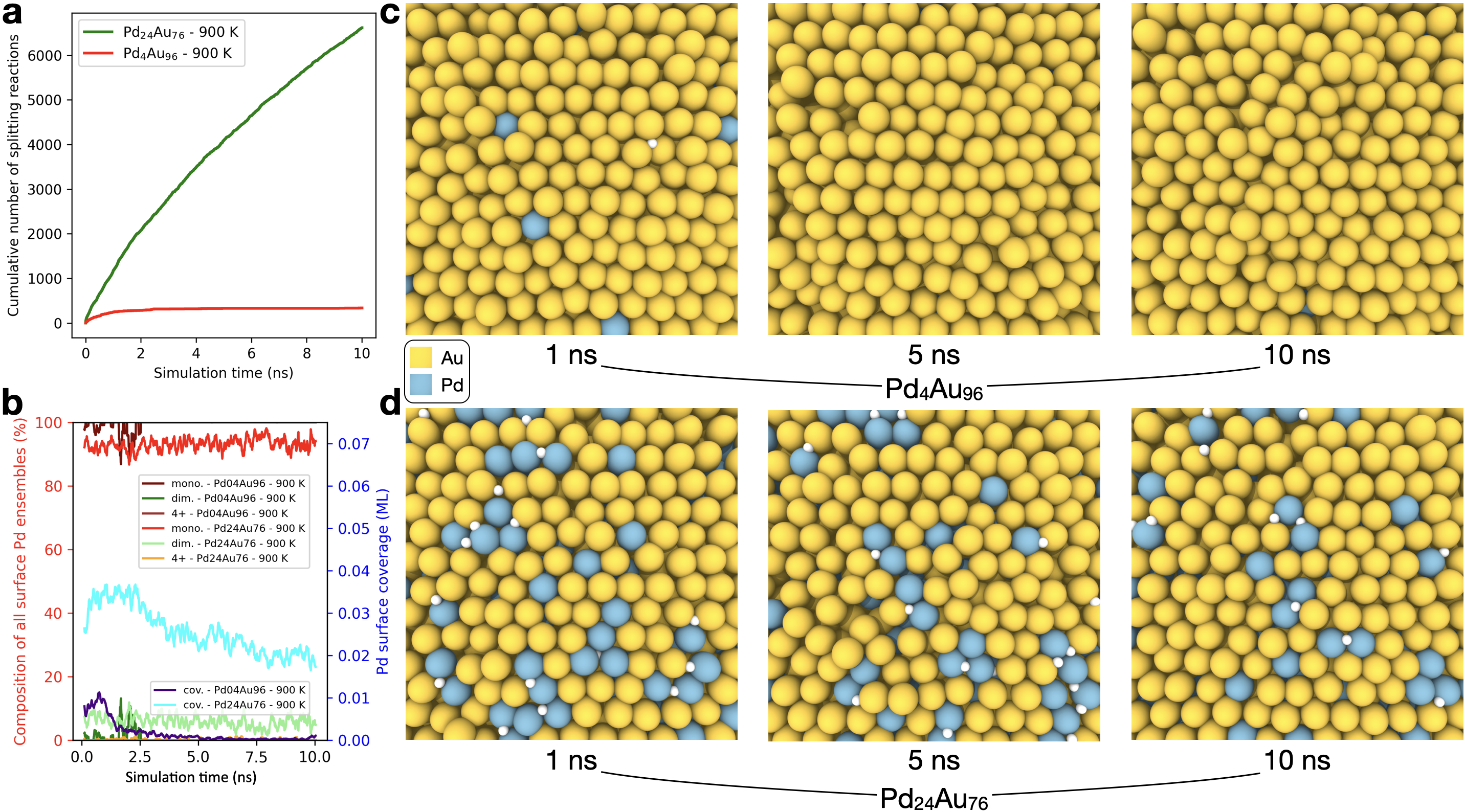}
\caption*{\textbf{Figure 7: Catalyst deactivation and active site ensemble evolution captured directly using simulation.} \textbf{a} Reactivity measurements for two separate nano-cells of the same size but different Pd content. The system with a composition of Pd$_4$Au${96}$ (red trace) quickly deactivates due to the well-established Pd bulk-segregation mechanism, while the catalyst with higher Pd content (composition of Pd$_{24}$Au${76}$) is much more active and more resistant to deactivation. \textbf{b} Evolution of Pd ensemble surface composition and surface coverage for Pd$_24$ and Pd$_04$ for the same systems presented in \textbf{a}. \textbf{c} Evolution of the Pd$_4$Au${96}$ catalyst across 10 ns of simulation time viewed from surface normal to the catalyst surface. Gold (yellow), palladium (blue), and hydrogen (white) can be readily observed. Hydrogen not bound to the surface is removed for clarity. The total amount of Pd is readily apparent across these simulation snapshots, directly explaining the loss of activity shown in \textbf{a}. \textbf{c} The same simulation snapshot series as in \textbf{b}, but for the Pd$_{24}$Au${76}$ catalyst. The overall length-scale of the Pd-ensembles decrease over simulation time, and approach a steady state distribution of small ensembles (e.g. monomers, dimers, and trimers).}
\label{fig:react1}
\end{figure*}

The high temperature data where catalyst deactivation is observed were removed from the Arrhenius analyses. The resulting fits on the reactivity data could then be used in a plot like Fig. 5d, where the pre-exponential factor and activation can be determined for the reaction on each catalyst. The complete results, provided in Table 2, show that the reactivity depends on the Pd ensemble size and Pd concentration. Pd monomers exhibit the lowest activation energy, consistent with previous predictions of Marcella et al. in Ref. \cite{Marcella2022}. The activation energy then increases with the Pd-ensemble size, as is shown by the dimer, trimer, and larger ensemble values, as well as those for Pd$_{8}$Au$_{92}$ and Pd$_{24}$Au$_{76}$. As the Pd concentration increases, the activation energies approach the pure Pd value of $\approx$ 0.45 eV, which is also predicted by our nanoreactor simulations containing a pure Pd slab, that provides and activation energy of 0.524 eV. 

\subsection*{Subtleties of adsorption reliable captured}
Important to the reactivity of this bimetallic system is the adsorption behavior of \ch{H2} on the various ensemble sizes as a function of temperature. Furthermore, the reaction profiles of Fig. 3 suggest that physisorption may be a dominant phenomenon, especially for systems comprised of Pd monomers at low temperatures. To answer this question of whether or not the subtleties of adsorption were captured, we computed various properties on the fly in the LAMMPS code, including H-H coordination, as well as H-Pd coordination. From the combination of these values, we could detect physisorption as well as dissociative chemisorption, the latter which was discussed in the previous section. These physisorption events are provided in Fig. 6, where H$_2$ is shown to physisorb on Pd-monomers as long as the simulation temperature is low. This is observed in panels \textbf{a} and \textbf{c}, where catalysts of very different composition exhibit the same phenomena, where isolated monomers lead to physisorption of the \ch{H2} molecule at low temperatures. At higher temperatures ($>$ 300 K), physisorption is no longer observed, as seen in panels \textbf{b} and \textbf{d}. These observations can be viewed in the context of the energy profiles provided in Fig. 3, where Pd-monomers on the Au(111) surface provide shallow energy minima for physisorption of H$_2$, and a sizable barrier for splitting. Hence, we can provide direct insight into the actual mechanism of \ch{H2} activation in our simulations as a function of the catalyst conditions and length-scale of the active site visited by the reactive adsorbate.

\subsection*{\textit{In silico} simulations predict catalyst deactivation}
Lastly, we studied the phenomenon of catalyst deactivation, as was noticed during the initial reactivity and Arrhenius analyses of the systems in prior sections. Catalyst deactivation is an impressively complex field of study, with various coupled mechanisms potentially responsible for the loss of activity within a catalytic material \cite{Martin2022}, so shifting responsibility and predictive ability to simulations would offer an incredible advantage for the maintenance and design of better next-generation catalysts for industrial use. Ultimately, it is known for Pd-Au catalysts that there is a strong energetic preference ($\approx$ 0.3 eV/atom) for Pd atoms to segregate into the bulk of Au, maximizing its partial coordination to Au, rather than remaining on the surface of the catalyst. Hence, we set out to determine the time-scales and mechanisms underlying the deactivation profiles in the reactivity data, and provide this analysis in Fig. 7. 

In panel \textbf{a}, we provide two high-temperature reactivity profiles for catalysts of the same size, but different Pd compositions. Focusing first on the red trace, we see that the catalyst quickly deactivates within the first 5 ns of simulation, as indicated by the horizontal profile of the splitting reactions. However, when the Pd concentrations is increased by a factor of 6, activity persists across a much longer simulation time. To explain this, we analyze the surface composition evolution, illustrated in panels \textbf{b} and \textbf{c}, for the Pd$_{4}$Au${96}$ and Pd$_{24}$Au${76}$ systems, respectively. In panel \textbf{b}, we see that no Pd is present on the catalyst surface after 5 ns, which directly explains the catalyst deactivation observed in panel \textbf{a}. In panel \textbf{c}, we see evolution of the Pd$_{24}$Au${76}$ catalyst, where the average size of the Pd ensembles decreases over the simulation time. Despite starting with large ensembles (e.g. Pd-tetramers and larger), the system approaches a steady-state composition that primarily consists of Pd-monomers, dimers and trimers. There is a coupled effect on the reactivity as a result of the decreased size of the Pd-ensembles, since their reduced sizes decreases the relative activation energy for hydrogen exchange, but the total amount of Pd is reduced, so this explains the slowing down of the reactivity with time, as shown in panel \textbf{a} (blue curve). In summary, our simulations provide the ability to directly study and design catalysts that are more resistant to deactivation mechanisms.

\subsection*{Discussion}
Ultimately, using the Allegro MLFF we were able to unveil the structural evolution of PdAu single crystals and NPs with and without reactive atmospheres and annealing conditions at quantum mechanical accuracy and experimentally relevant time- and length-scales. These insights demonstrate the subtleties of adsorption on these small length-scale Pd active sites on hydrogen activation, where active sites and their evolution can be determined with relative ease under realistic conditions, and mechanisms of critical processes can be discovered and explained atomistically, including hydrogen spillover to the less active Au surface, and the preference of Pd to split into smaller ensemble sizes as well as diving into the subsurface to increase coordination with Au, reducing the surface energy of the system simultaneously. These observations were enabled via the ability of this combination of methods to yield a single, comprehensive model that can encompass the geometric and chemical complexity associated with heterogeneous catalysts, and directly, and simultaneously, model catalyst degradation and explicit reactivity. This opens the door towards simultaneous study of alloying and reaction mechanisms in tandem, and these observations will inform catalyst design. Supporting these claims is the fact that our MLFF is able to verify previous findings using traditional DFT and experimental techniques, including (1) evolved NP symmetries and redistribution of the active Pd ensembles across concentrations and NP sizes, (2) available Pd-ensembles for reactivity and their stability under such stimuli, and (3) simultaneous observation of catalyst degradation mechanisms while accounting for explicit reactivity. 

This work establishes a workflow from which bimetallic NPs and surfaces can be understood with atomistic resolution at quantum mechanical accuracy under realistic conditions, since this simulation protocol can be trivially extended to systems with different compositions and reactive atmospheres. Critical to this extension is the computational workflow employed, where highly efficient parallel active learning trajectories are used to generate training data for the construction of an accurate and reliable equivariant Allegro MLFF that can be used to evolve in silico twins of experimental systems. Ultimately,this method is employed to study active site evolution as a function of alloy composition, NP size, and adsorbate exposure, providing atomic-level heuristics by which practitioners can begin to design better catalysts that maintain high activity and selectivity while being resistant to degradation mechanisms. 

\subsection*{Methods}
\subsection*{FLARE active learning allows for efficient collection of first-principles training data}
We find it pertinent to provide a brief discussion of the FLARE and Allegro training processes for this complex catalytic system. This pipeline is shown in Fig. 1, where FLARE feeds data to Allegro, the resulting model of which is then used to study the dynamic evolution of PdAu catalysts under a variety of conditions.  In short, a variety of small, DFT-accessible atomic structures were used as input for the active learning routine of FLARE. Several of these active learning trajectories were then run in parallel on a large cluster of CPU nodes, all employing DFT at the PBE level. These parallel active learning trajectories are summarized in Suppl. Table 1. Ultimately, a total of 20,816 DFT calculations were performed on relevant systems containing various structural motifs, system sizes, and reactive chemistries for H-Pd-Au within a maximum individual walltime of 168 hours when all run in parallel. This data set was then combined with the 2965 frames from \cite{Owen2024_au} for the description of Au bulk, surfaces, and NPs. If run in serial without `warm-starts' using a pre-trained FLARE model for each subsequent trajectory, the data set would take 425.6 days to collect, but this is negated by running the trajectories in parallel. This complete set of data was then fed into the Allegro architecture to train the final MLFF used for all production ML-MD simulations. These methods, both FLARE and Allegro are described in more detail below.

\subsection*{Bayesian Active-Learning in FLARE}
The Fast Learning of Atomistic Rare Events (FLARE) open-source code: \url{https://github.com/mir-group/flare}, was employed to efficiently collect \textit{ab initio} cells for H-Pd-Au systems. Briefly, the spatial geometry of local atomic environments (within a cutoff radius $r_c$) was encoded in descriptors using the atomic cluster expansion (ACE) \cite{Drautz2019AtomicPotentials}. The normalized dot-product kernel was used to measure the similarity between atomic descriptors, which was then used in the construction of a sparse Gaussian process regression model (SGP) to predict the atomic energies of such local environments. The SGP surrogate model provides an inherent mechanism to quantify predictive uncertainties for atomic forces, energies, and stresses, which can be used within the active learning algorithm to select \textit{ab initio} training data `on-the-fly' during the MD simulation. After training is finished, following an \textit{ab initio} calculation of a given frame, the SGP force field was then remapped and the surrogate was used to continue the MD simulation within the active-learning loop. 

The ACE-B2 descriptor was employed for all parallel active-learning trajectories. By using the second power of the normalized dot product kernel, `effective' 5-body interactions within each descriptor were obtained, which is sufficiently complex for describing Au and Pd with high accuracy \cite{owen2024complexity}. Maintaining consistency in notation with the original work of ACE \cite{Drautz2019AtomicPotentials}, ultimately, we chose $n_{\text{max}}=8$ for radial basis, $l_{\text{max}}=3$ for angular basis, and the cutoff radius $r_{cut} = 6 \rm{\AA}$ for all Au-Pd, Au-H, Pd-H, and the inverse of such interactions. We refer the reader to Ref. \cite{Vandermause2022} for more in-depth discussion of this process and the background mathematical arguments.

\subsection*{Density Functional Theory}
All plane-wave DFT calculations were performed in the Vienna \textit{Ab Initio} Simulation Package (VASP, v5.4.4) using the Perdew-Burke-Ernzerhof exchange correlation functional in the Projector Augmented Wave (PAW) formalism. The Au, Pd, and H pseudopotentials each contained 11, 10, and 1 valence electrons, respectively. Electronic smearing within the Methfessel-Paxton scheme \cite{PhysRevB.40.3616} (ISMEAR = 1 in VASP) was set to 0.2 eV for all calculations. The \textbf{k}-point spacing was chosen such that the energy noise per atom was below 1 meV/atom and the force noise was below 5 meV/\AA{}, which corresponds to a maximum \textbf{k}-spacing of 0.2 \AA{}$^{-1}$. All \textbf{k}-point grids were centered at the gamma-point, and only the gamma-point was employed for along non-periodic directions for slabs, nanoparticles, and gaseous cells. A cutoff energy of 450 eV was employed, and the cutoff energy of the augmentation charges was set to 1800 eV. Spin-polarization was not included for all calculations. An energy threshold of 1E-05 was employed for static calculations, and a force convergence threshold of -1E-03 was employed for all ionic relaxations. For the reaction pathways provided in Fig. 3 of the Main text, structures were taken from the SI of Ref. \cite{Marcella2022} and recomputed with the same level of DFT that was used to construct the MLFF training set. The methods used to determine the transition states are described in \cite{Marcella2022}.

\subsection*{Allegro MLFF Training}
We then trained an equivariant Allegro MLFF on the complete set of first-principles data collected using FLARE active learning as described in the previous section. During this process, we explicitly tested the effect of varying model architectures on a wide-ranging set of validation targets. Ultimately, a model with an angular resolution of $\ell_{max}=2$ and symmetric pairwise cutoff matrix with values 6.0, 5.0, and 5.0 \AA{} was employed for the Au-Au/Pd-Pd, H-Au/H-Pd, and H-H interactions, respectively. These parameters appeared to be most critical for this system, where the choice of a larger cutoff for the H-H interaction in the larger cutoff matrix was found to be necessary for a more accurate description of bulk Pd-hydride, since smaller cutoffs resulted in a drastic shift of the predicted minimum in the bulk energy as a function of volume to lower, nonphysical values. Following this fairly exhaustive grid search over model parameters, with a held out test set being the critical criterion for model `success,' the final network employed two layers, 8 tensor features, multi-layer perceptron input dimensions of [64,128,256] for the scalar track, latent dimensions of [256,256,256], angular resolution of $\ell_{max}=2$, with an interaction cutoff of 6 \AA{}, and 5 \AA{} for H-M interactions. This model architecture was chosen based on the training and validation errors, as they were tracked using weights and biases, as well as model performance on the held-out bulk, surface, NP, and adsorbate test set discussed in the Main text. The model used a `default\_dtype` of 64 for floating point operations, and the final model was deployed with tensor float 32 turned off. Lastly, the loss coefficients for energies, forces, and stresses were set to 25, 1, and 1, respectively, where the energy loss coefficient employed the PerAtomMSELoss option.

\subsection*{Molecular Dynamics Simulations}
All MD simulations were performed in LAMMPS \cite{THOMPSON2022108171}, where a custom Allegro pairstyle was employed \cite{10.1145/3581784.3627041}. GPU acceleration was achieved with the Kokkos portability library \cite{CARTEREDWARDS20143202}. All simulations were done in the Nos\'e-Hoover NVT ensemble. A timestep of 5 femtoseconds was employed for all simulations without hydrogen, while a timestep of 0.2 femtoseconds was employed for all simulations with hydrogen. Velocities were randomly initialized for all simulations to a Boltzmann distribution centered at whatever desired temperature for the simulation (in units of Kelvin). All surfaces were constructed using the minimized lattice constant of Au (4.16 \AA{}) in the Atomic Simulation Environment (ASE) \cite{Larsen_2017}, and Pd was introduced using atom-type substitutions in LAMMPS. For reactive simulations, H$_2$ was introduced into the cell using ASE. Velocity rescaling was performed at a frequency of 100$\times$dt for all simulations, and angular and linear momentum were rescaled for Au and Pd atoms every timestep. The exact scripts used for the simulations will be provided in the Materials Cloud Archive repository.

\subsection*{Trajectory Analysis}
Each LAMMPS simulation was dumped using the LAMMPS binary format to conserve memory. A custom LAMMPS binary `dump-reader' was employed (\url{https://github.com/anjohan/lammps-binary-dump-reader}) to extract thermodynamic results from the simulations. Ovito python scripts were created to compute reaction counts, and total ensemble counts as a function of simulation time. These scripts will be provided in the Materials Cloud Archive upon publication. For determination of the Pd ensemble sizes throughout each simulation, coordination numbers were computed on-the-fly during each lammps trajectory, and then comfirmed using the Ovito script \cite{ovito}. Following parsing of the simulation data, analysis, and recording, python scripts employing matplotlib.pyplot were used to plot the results.

\subsection*{Ensemble and Coordination Analysis}
The NP simulation protocol consisted of annealing at constant temperature for 10 ns (with temperatures ranging from 500 - 760K based on NP size and alloy composition), followed by quenching to 300K at a rate of 20 K/ns. For each size and composition of alloy nanoparticles, trajectories were run in three sets of identical simulations, each differing solely in the initial configuration, where we began with a full Pd structure and substituted Au atoms randomly. This variation was achieved using different random seeds set within the LAMMPS script to place the Au and Pd atoms within the nanoparticle.

Both ensemble and partial coordination numbers were analyzed using the Ovito python API \cite{ovito}. We computed p(CN) with a cutoff radius of 3.5 \AA{} for the pairs Pd-Au, Au-Au, Au-Pd, Pd-Pd in the final quenched structure and averaged across the three replicate trajectories, and plotted information relative to the first pair type (Pd-Au) in Fig. 4a. We employed the Ovito Cluster Analysis modifier to identify Pd surface ensembles with a cutoff of 3 \AA{}. The time evolution of ensembles information and Pd coverage of the surface was collected for each trajectory and averaged across the three replicates. To minimize the influence of the initial random placements on the data, the first nanosecond of each trajectory was omitted from the ensemble analysis. Finally, the time evolution information  was plotted as a 1 ns trailing average for both Pd surface coverage, and monomer percentage of surface Pd ensembles, to highlight the long trends over short term variability.

\subsection*{Arrhenius Analysis}
In order to extract the activation energy and pre-exponential factor for hydrogen exchange over each catalyst, Arrhenius analyses were performed for each system. This was done using the OVITO python API \cite{ovito}, where the reactivity data for both dissociative chemisorption and recombinative desorption of H$_2$ was fed into the pipeline and plotted as a function of simulation time, as is shown in Fig. 5c and Fig. 7a of the Main text. These data were then fit using linear regression over the last 1 ns of simulation time, from which the slope was extracted and the natural logarithm was performed. These values, extracted at each temperature, where then plotted as a function of inverse temperature, where another linear regression was performed to extract both the slope (activation energy) and the intercept (pre-exponential factor).

\subsection*{Data Availability}
The HPdAu Allegro MLFF, \textit{ab initio} training data, and simulation input scripts will be provided on the Materials Cloud Archive upon publication.

\subsection*{Code Availability}
The details about VASP, a proprietary code, can be found at https://www.vasp.at/. The details about FLARE and Allego, which are open-source codes, can be found at https://github.com/mir-group/flare and https://github.com/mir-group/allegro, respectively.

\subsection*{Author Contributions} 
C.J.O. created the data set using FLARE active-learning, compiled and analyzed the data, performed all Allegro MLFF training, validation, MD simulations, and wrote the manuscript. L.R. completed the NP alloying simulations and analysis that is provided in Fig. 4 of the main text. C.R.O. and N.M. provided helpful discussions regarding the experimental interpretation of the results. A.M. provided helpful discussions pertaining to the Allegro method during training. B.K. supervised all aspects of the work. All authors contributed to revision of the manuscript.

\subsection*{Acknowledgements} 
We acknowledge Jin Soo Lim and Lixin Sun for helpful discussions at the outset of this work. This work was supported by the US Department of Energy, Office of Basic Energy Sciences Award No. DE-SC0022199 and No. DE-SC0020128, as well as by Robert Bosch LLC. An award for computer time was provided by the U.S. Department of Energy’s (DOE) Innovative and Novel Computational Impact on Theory and Experiment (INCITE) Program. This research used supporting resources at the Argonne and the Oak Ridge Leadership Computing Facilities. The Argonne Leadership Computing Facility at Argonne National Laboratory is supported by the Office of Science of the U.S. DOE under Contract No. DE-AC02-06CH11357. The Oak Ridge Leadership Computing Facility at the Oak Ridge National Laboratory is supported by the Office of Science of the U.S. DOE under Contract No. DE-AC05-00OR22725. Additional computational resources were provided by the FAS Division of Science Research Computing Group at Harvard University and resources of the National Energy Research Scientific Computing Center (NERSC), a DOE Office of Science User Facility supported by the Office of Science of the U.S. Department of Energy under Contract No. DE-AC02-05CH11231 using NERSC award BES-ERCAP0024206.

\subsection*{Competing interests}
The authors declare no competing interests.

\subsection*{References}
\bibliography{bib}

\end{document}

%\title{Atomistic Evolution of PdAu Catalysts \\Across Compositions and Chemistries}
\title{Supplementary Information for: Atomistic evolution of active-sites in multi-component heterogeneous catalysts}

\author{Cameron J. Owen$^{\dagger}$}
\affiliation{Department of Chemistry and Chemical Biology, Harvard University, Cambridge, Massachusetts 02138, United States}
\affiliation{John A. Paulson School of Engineering and Applied Sciences, Harvard University, Cambridge, Massachusetts 02138, United States}

\author{Lorenzo Russotto}
\affiliation{Department of Physics, Harvard University, Cambridge, Massachusetts 02138, United States}

\author{Christopher R. O'Connor}
\affiliation{Rowland Institute at Harvard, Harvard University, Cambridge, Massachusetts 02142, United States}

\author{Nicholas Marcella}
\affiliation{Department of Chemistry, University of Illinois, Urbana, Illinois 61801, United States}

\author{Anders Johansson}
\affiliation{John A. Paulson School of Engineering and Applied Sciences, Harvard University, Cambridge, Massachusetts 02138, United States}

\author{Albert Musaelian}
\affiliation{John A. Paulson School of Engineering and Applied Sciences, Harvard University, Cambridge, Massachusetts 02138, United States}

\author{Boris Kozinsky$^{\dagger}$}
\affiliation{John A. Paulson School of Engineering and Applied Sciences, Harvard University, Cambridge, Massachusetts 02138, United States}
\affiliation{Robert Bosch LLC Research and Technology Center, Watertown, Massachusetts 02472, United States}

\def\thefootnote{$\dagger$}\footnotetext{Corresponding authors\\C.J.O., E-mail: \url{cowen@g.harvard.edu}\\B.K., E-mail: \url{bkoz@seas.harvard.edu}\\ }\def\thefootnote{\arabic{footnote}}

% custom commands
\newcommand\bvec{\mathbf}
\newcommand{\mathsc}[1]{{\normalfont\textsc{#1}}}

\maketitle

\begin{table*}[!htbp]
\centering
\resizebox{\textwidth}{!}{\begin{tabular}{|c|c|c|c|c|c|c|c|c|}
\multicolumn{1}{c}{\bf System}    & \multicolumn{1}{c}{\bf Ensemble} & \multicolumn{1}{c}{\bf Temp. (K)} & \multicolumn{1}{c}{\bf Pres. (GPa)}   &  \multicolumn{1}{c}{\bf $\sum\tau_{\textrm{sim}}$ (ns)} & \multicolumn{1}{c}{\bf   $\sum\tau_{\textrm{wall}}$ (hr)} & \multicolumn{1}{c}{\bf $\sum N_{\textrm{DFT}}$} & \multicolumn{1}{c}{\bf $\sum N_{\textrm{runs}}$} & \multicolumn{1}{c}{\bf $\sum N_{\textrm{atoms}}$}\\
\hline
\hline
Au-Recon. \cite{Owen2024_au} & NVT & 250-1500 & --- & 13.2 & 1775.1 & 2965 & 46 & 55-309 \\ 
\hline
\multirow{2}{*}{Au$_{bulk,bcc}$} & \multirow{2}{*}{NPT-iso.} & \multirow{2}{*}{300-1500} & -1.0 & 4.79 & 36.8 & 17 & 1 & \multirow{2}{*}{54} \\
                &          &  & 1.0 & 9.29 & 25.3 & 11 & 1 &  \\
\hline
\multirow{2}{*}{Au$_{bulk,fcc}$} & \multirow{2}{*}{NPT-iso.} & \multirow{2}{*}{300-1500} & -1.0 & 2.77 & 17.9 & 17 & 1 & \multirow{2}{*}{32} \\
                &          &  & 1.0 & 3.09 & 8.9 & 8 & 1 &  \\
\hline
\multirow{2}{*}{Au$_{bulk,hcp}$} & \multirow{2}{*}{NPT-iso.} & \multirow{2}{*}{300-1500} & -1.0 & 0.92 & 15.1 & 9 & 1 & \multirow{2}{*}{32} \\
                &          &  & 1.0 & 1.91 & 12.0 & 7 & 1 &  \\
\hline
\multirow{2}{*}{Pd$_{bulk,bcc}$} & \multirow{2}{*}{NPT-iso.} & \multirow{2}{*}{300-2000} & -1.0 & 7.35 & 19.4 & 6 & 1 & \multirow{2}{*}{54} \\
                &          &  & 1.0 & 0.21 & 10.7 & 4 & 1 &  \\
\hline
\multirow{2}{*}{Pd$_{bulk,fcc}$} & \multirow{2}{*}{NPT-iso.} & \multirow{2}{*}{300-2000} & -1.0 & 1.16 & 6.3 & 6 & 1 & \multirow{2}{*}{32} \\
                &          &  & 1.0 & 2.92 & 4.9 & 5 & 1 &  \\
\hline
\multirow{2}{*}{Pd$_{bulk,hcc}$} & \multirow{2}{*}{NPT-iso.} & \multirow{2}{*}{300-2000} & -1.0 & 2.52 & 10.9 & 5 & 1 & \multirow{2}{*}{32} \\
                &          &  & 1.0 & 1.94 & 11.4 & 6 & 1 &  \\
\hline
\multirow{2}{*}{PdAu$_{bulk}$} & \multirow{2}{*}{NPT-iso.} & \multirow{2}{*}{300-2000} & -1.0 & 10.0 & 131.3 & 315 & 1 & \multirow{2}{*}{16} \\
                &          &  & 1.0 & 10.0 & 100.3 & 241 & 1 &  \\
\hline
\multirow{2}{*}{PdAu$_{3,bulk}$} & \multirow{2}{*}{NPT-iso.} & \multirow{2}{*}{300-2000} & -1.0 & 10.0 & 48.1 & 62 & 1 & \multirow{2}{*}{32} \\
                &          &  & 1.0 & 10.0 & 38.6 & 53 & 1 &  \\
\hline
\multirow{2}{*}{Pd$_3$Au$_{5,bulk}$} & \multirow{2}{*}{NPT-iso.} & \multirow{2}{*}{300-2000} & -1.0 & 10.0 & 94.8 & 107 & 1 & \multirow{2}{*}{32} \\
                &          &  & 1.0 & 10.0 & 51.6 & 65 & 1 &  \\
\hline
\multirow{2}{*}{Pd$_3$Au$_{bulk}$} & \multirow{2}{*}{NPT-iso.} & \multirow{2}{*}{300-2000} & -1.0 & 10.0 & 34.1 & 39 & 1 & \multirow{2}{*}{32} \\
                &          &  & 1.0 & 10.0 & 23.9 & 30 & 1 &  \\
\hline
\multirow{2}{*}{Pd-H$_{bulk,convex-hull}$} & \multirow{4}{*}{NVT} & \multirow{4}{*}{300-2000} & -1.0 & 0.71 & 166.9 & 130 & 1 & 64 \\
 &  &  & 1.0 & 0.90 & 165.8 & 133 & 1 & 64 \\
\multirow{2}{*}{Pd-H$_{bulk,high}$} &  &  & -1.0 & 6.68 & 167.4 & 224 & 1 & 32 \\
&  &  & 1.0 & 9.37 & 161.3 & 204 & 1 & 32 \\
\hline
Au(100) & \multirow{3}{*}{NVT} & \multirow{3}{*}{300-2000} & \multirow{3}{*}{---} & 10.0 & 59.4 & 72 & 1 & 54 \\
Au(110) &  &  &  & 10.0 & 86.3 & 74 & 1 & 60 \\
Au(111) &  &  &  & 10.0 & 88.6 & 85 & 1 & 54 \\
\hline
Pd(100) & \multirow{3}{*}{NVT} & \multirow{3}{*}{300-2000} & \multirow{3}{*}{---} & 10.0 & 66.9 & 51 & 1 & 54 \\
Pd(110) &  &  &  & 10.0 & 88.3 & 48 & 1 & 60 \\
Pd(111) &  &  &  & 10.0 & 43.0 & 34 & 1 & 54 \\
\hline
Au$_{55,ico.}$ & \multirow{3}{*}{NVT} & 400-800 & \multirow{3}{*}{---} & 2.80 & 78.6 & 392 & 6 & 55 \\
Au$_{147,ico.}$ &  & 500-1200 &  & 2.40 & 46.2 & 74 & 3 & 147 \\
Au$_{309,ico.}$ &  & 500-800 &  & 1.11 & 145.2 & 106 & 4 & 309 \\
\hline
Pd$_{55,ico.}$ & \multirow{5}{*}{NVT} & \multirow{5}{*}{300-2000} & \multirow{5}{*}{---} & 1.77 & 315.2 & 136 & 3 & 55 \\
Pd$_{55,cub.}$ &  &  &  & 0.47 & 282.3 & 122 & 3 & 55 \\
Pd$_{85,oct.}$ &  &  &  & 5.24 & 295.2 & 159 & 3 & 85 \\
Pd$_{147,ico.}$ &  &  &  & 0.07 & 286.4 & 84 & 3 & 147 \\
Pd$_{147,cub.}$ &  &  &  & 1.87 & 256.6 & 105 & 2 & 147 \\
\hline
Au$_{13}$Pd$_{42,core-shell,ico.}$ & \multirow{9}{*}{NVT} & 800 & \multirow{9}{*}{---} & 1.00 & 33.0 & 127 & 1 & 55 \\
Au$_{13}$Pd$_{42,ran.-alloy,ico.}$ &  & 800 &  & 1.00 & 32.6 & 151 & 1 & 55 \\
Au$_{55}$Pd$_{92,core-shell,ico.}$ &  & 900 &  & 1.00 & 80.4 & 87 & 1 & 147 \\
Au$_{13}$Pd$_{134,core-2\cdot shell,ico.}$ &  & 900 &  & 1.00 & 41.5 & 149 & 1 & 147 \\
Au$_{55}$Pd$_{92,ran.-alloy,ico.}$ &  & 900 &  & 1.00 & 59.8 & 56 & 1 & 147 \\
Au$_{13}$Pd$_{134,ran.-alloy,ico.}$ &  & 900 &  & 1.00 & 48.3 & 57 & 1 & 147 \\
Au$_{147}$Pd$_{162,core-shell,ico.}$ &  & 900 &  & 0.05 & 168.9 & 96 & 3 & 309 \\
Au$_{55}$Pd$_{254,core-2\cdot shell,ico.}$ &  & 900 &  & 0.02 & 38.86 & 27 & 1 & 309 \\
Au$_{55}$Pd$_{254,ran.-alloy,ico.}$ &  & 800-900 &  & 0.40 & 59.99 & 58 & 2 & 309 \\
\hline
H/Au(100) & \multirow{3}{*}{NVT} & \multirow{3}{*}{300-2000} & \multirow{3}{*}{---} & 0.009 & 259.4 & 569 & 3 & 90 \\
H/Au(110) &  &  &  & 0.008 & 259.3 & 509 & 3 & 84 \\
H/Au(111) &  &  &  & 0.005 & 268.5 & 509 & 3 & 90 \\
\hline

H/Pd(100) & \multirow{3}{*}{NVT} & \multirow{3}{*}{300-2000} & \multirow{3}{*}{---} & 0.032 & 345.2 & 658 & 4 & 90-108 \\
H/Pd(110) &  &  &  & 0.024 & 274.3 & 548 & 4 & 84-128 \\
H/Pd(111) &  &  &  & 0.051 & 352.4 & 525 & 4 & 90-108 \\
\hline

H/Pd$_{0.20\text{ML}}$/Au(111) & \multirow{5}{*}{NVT} & \multirow{5}{*}{300-2000} & \multirow{5}{*}{---} & 0.006 & 365.2 & 1077 & 4 & 150 \\
H/Pd$_{0.36\text{ML}}$/Au(111) &  &  &  & 0.007 & 305.3 & 931 & 4 & 150 \\
H/Pd$_{0.56\text{ML}}$/Au(111) &  &  &  & 0.003 & 281.7 & 861 & 4 & 150 \\
H/Pd$_{0.80\text{ML}}$/Au(111) &  &  &  & 0.005 & 291.9 & 831 & 4 & 150 \\
H/Pd$_{1.00\text{ML}}$/Au(111) &  &  &  & 0.004 & 292.1 & 851 & 4 & 150 \\

\hline
H/Au$_{55,ico.}$ & \multirow{6}{*}{NVT} & \multirow{6}{*}{300-2000} & \multirow{6}{*}{---} & 0.153 & 150.3 & 518 & 4 & 147 \\
H/Au$_{147,ico.}$ &  &  &  & 0.002 & 12.8 & 44 & 1 & 309 \\
H/Au$_{13}$Pd$_{42,core-shell,ico.}$ &  &  &  & 0.015 & 123.4 & 421 & 3 & 147 \\
H/Au$_{1}$Pd$_{54,core-2\cdot shell,ico.}$ &  &  &  & 0.083 & 220.7 & 611 & 4 & 147 \\
H/Au$_{1}$Pd$_{54,ran.-alloy,ico.}$ &  &  &  & 0.025 & 232.6 & 549 & 4 & 147 \\
H/Au$_{55}$Pd$_{92,core-shell,ico.}$ &  &  &  & 0.0002 & 6.3 & 21 & 1 & 309 \\
\hline

Pd-H$_{\text{NP},98\text{ atoms},convex-hull,min.}$ & \multirow{3}{*}{NVT} & \multirow{3}{*}{300-2000} & \multirow{3}{*}{---} & 0.020 & 213.1 & 476 & 2 & 98 \\
Pd-H$_{\text{NP},159\text{ atoms},convex-hull,min.}$ &  &  &  & 0.013 & 83.3 & 149 & 1 & 159 \\
Pd-H$_{\text{NP},216\text{ atoms},convex-hull,min.}$ &  &  &  & 0.0002 & 35.7 & 104 & 2 & 216 \\

\hline
\hline
\textbf{Total} & \textbf{---} & \textbf{---} & \textbf{---} & 232.7 & 10213.7 & 23781 & 178 & \textbf{---} \\
\hline
\end{tabular}}
\caption*{\textbf{Table 1: Summary of the FLARE active learning procedure for H-Pd-Au training set construction.} The values provided for each system represent the sum across several independent trajectories run in parallel, the total number of which is provided in the second-to-last column.}
\label{table:active}
\end{table*}

\begin{figure*}
\centering
\includegraphics[width=\textwidth]{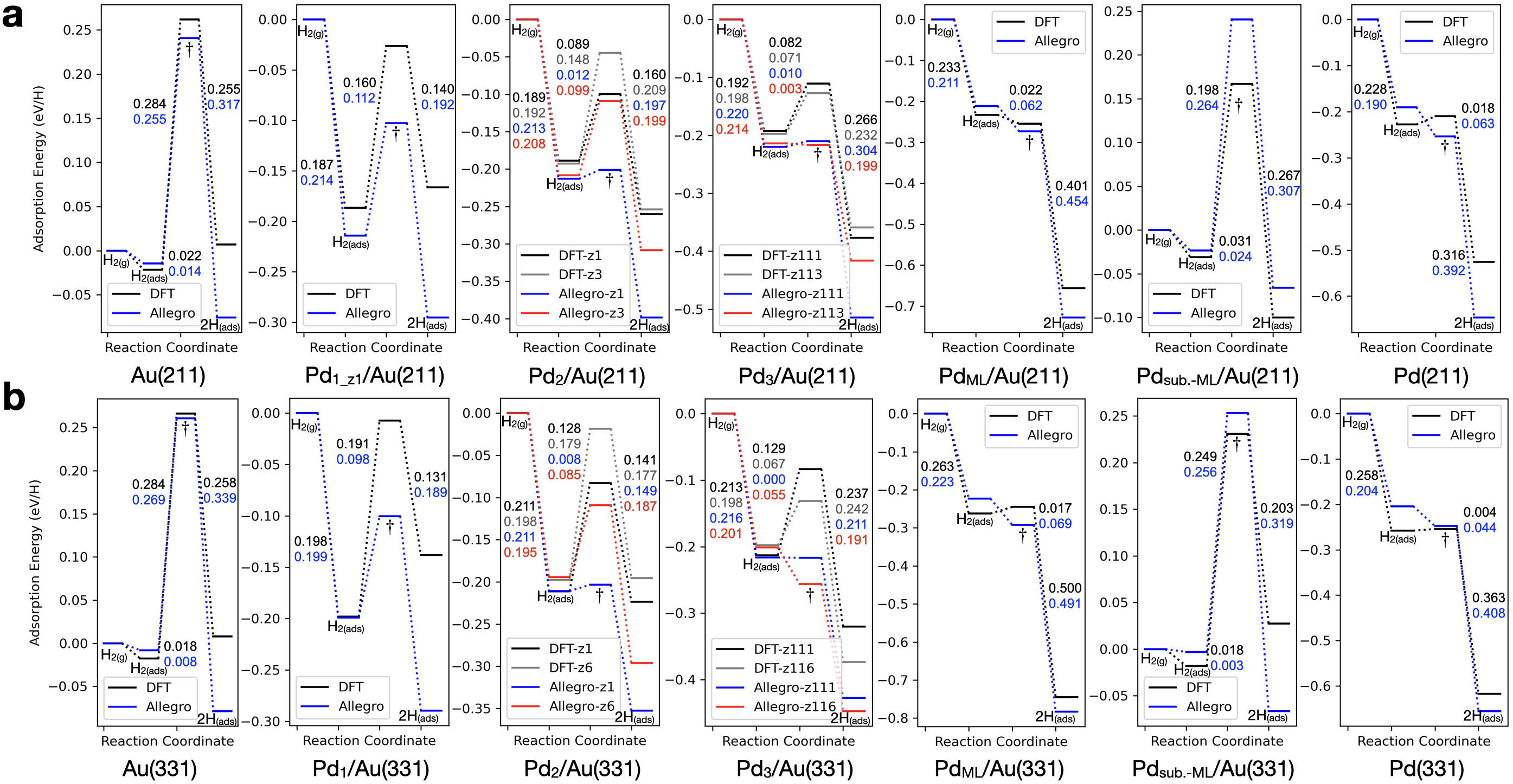}
\caption*{\textbf{Supplementary Figure 1: Additional adsorption validation of the MLFF against DFT.} \textbf{a} H$_2$ physisorption $\to$ 2H chemisorption reaction barriers for pure Au(211), dilute Pd ensembles in Au(211), and Pd(211) predicted by DFT (black and gray) and Allegro (blue and red). Transition states are identified by a cross. \textbf{b} H$_2$ physisorption $\to$ 2H chemisorption reaction barriers for pure Au(311), dilute Pd ensembles in Au(311), and Pd(311) predicted by DFT (black and gray) and Allegro (blue and red). Transition states are identified by a cross. Naming conventions for transition states follow the format defined in Ref. \cite{Marcella2022}.}
\label{fig:ads}
\end{figure*}

\subsection*{References}
\bibliography{bib.bib}